\documentclass[a4paper,fleqn,usenatbib, linenumber]{mnras}

\usepackage[T1]{fontenc}
\usepackage{ae,aecompl}
\usepackage[utf8]{inputenc}

\usepackage{graphicx}	% Including figure files
\usepackage{amsmath}	% Advanced maths commands
\usepackage{amssymb}	% Extra maths symbols
\usepackage{booktabs}
\title[]{Broadband study of OQ 334 during its flaring state}

\author[]{
Raj Prince$^{1}$\thanks{E-mail: raj@cft.edu.pl},
Rukaiya Khatoon$^{2}$, C. S. Stalin$^{3}$
\\
\\
$^{1}$Center for Theoretical Physics, Polish Academy of Sciences, Al.Lotnikov 32/46, 02-668, Warsaw, Poland\\
$^{2}$Tezpur University, Napaam-784028, Assam, India\\
$^{3}$Indian Institute of Astrophysics, Block II, Koramangala, Bangalore - 560034, India \\
}

\date{Accepted XXX. Received YYY; in original form ZZZ}

\begin{document}
\label{firstpage}
\pagerange{\pageref{firstpage}--\pageref{lastpage}}
\maketitle
\begin{abstract}

The blazar OQ 334 displayed a $\gamma$-ray flare in 2018, after being in the long quiescent $\gamma$-ray state since 2008. Subsequent to the flare, the source was in a higher $\gamma$-ray flux state and again flared in 2020. We present here the first spectral and timing analysis of the source at its various flaring states. During the higher $\gamma$-ray state, we found four major peaks identified as P1, P2, P3 and P4. From timing analysis we found rise and
decay time of the order of hours with the fastest variability time of 9.01 $\pm$ 0.78 hr. We found the highest $\gamma$-ray photon of 77 GeV during P4, which suggests the location of the $\gamma$-ray emitting region at the outer edge of the broad line region or the inner edge of the torus. The $\gamma$-ray spectral analysis of the source indicates that during P4, the $\gamma$-ray spectrum clearly deviates from the power law behaviour.
From cross-correlation analysis of the $\gamma$-ray and radio lightcurves, we found that the two emission regions are separated by about 11 pc. Our broad band spectral energy distribution modeling of the source during quiescent and active phases indicates that more electron and proton power are required to change the source from low flux to high flux state. 
The Anderson-Darling test and histogram fitting results suggest that the three days binned 
$\gamma$-ray fluxes follow a lognormal distribution.

%The $\gamma$-ray flux distribution from the source is found to have a log-normal pattern which argues for $\gamma$-ray flux variations due to  processes in the relativistic jet of the source.

\end{abstract}


%%%%%%%%%%%%%%%%%%%%%%%%%%%%%%%%%%%%%%%%%%%%%%%%%%

%%%%%%%%%%%%%%%%% BODY OF PAPER %%%%%%%%%%%%%%%%%%

\section{Introduction}
Blazars are a peculiar category of active galactic nuclei (AGN) that have their relativistic jets aligned close to the line of sight (angles less than $\sim$ 14 $^{\circ}$) to the observer (\citealt{Urry_1995}). Their energy output dominated by non-thermal emission spans the complete accessible electromagnetic spectrum. Blazars are highly luminous, have powerful jets, powered by massive black holes (\citealt{Lynden-Bell_1969}) and dominate the extragalactic $\gamma$-ray sky (\citealt{Hartman_1999}, \citealt{Abdo_2010}). They show large amplitude flux variability in different wavelengths such as radio, infrared, optical, X-rays and $\gamma$-rays on a range of time scales from minutes to hours to several days (\citealt{Heidt_1996}, \citealt{Ulrich_1997} ) and in some cases to decades (\citealt{Goyal_2017, Goyal_2018, Goyal_2020}); For a review see, \citealt{Hovatta_2019}.
The increased capability in the recent years to acquire near simultaneous observations over different wavelengths have led to the identification of complex variability patterns across wavelengths in blazars. There are instances when the variations in the low energy optical and high energy $\gamma$-rays are correlated and also instances where uncorrelated variations between optical and $\gamma$-rays are noticed( \citealt{Chatterjee_2012}, \citealt{Rajput_2019, Rajput_2020}). Also, the short time scale of variations now observed in blazars indicate that the variations we observe in them arise from very small regions in their jets. 

The broad band spectral energy distribution (SED) of blazars show two distinct humps. The low energy hump peaks in the UV/X-ray region and is now understood to be due to synchrotron emission from relativistic electrons in their jet. The high energy hump peaks in the MeV-GeV energy range and the physical mechanisms responsible for the high energy hump is still debated. In the widely used leptonic model of emission from blazar jets, the high energy emission is due to inverse Compton process. The seed photons for the inverse Compton process can be photons internal to the jet (synchrotron self Compton; \citealt{Sikora_2009}) or  external to the jet (external compton; \citealt{Dermer_92}; \citealt{Sikora_94}). Alternative to the leptonic process is the hadronic process. In this model, the high energy emission is explainable by proton
synchrotron process or photo-pion process (\citealt{Bottcher_2013}). Thus, carrying out timing and SED analysis of blazars can provide valuable clues to the processes happening close to the central regions of blazars. In spite of such studies done on blazars, we yet do not have a clear understanding of the physical processes happening in blazar jets. 

The observed broad band SED of blazars is complex. For example, SED modelling of 3C 279 using the one zone leptonic emission model during the flares
in 2017-2018, favours the $\gamma-ray$ emission site to be located at the outer boundary of BLR (\citealt{Prince_2020}). Also, the flare of 2014 March in 3C 279 was not detected in the Very high energy gamma-ray band, and the Fermi observations were explained in the one-zone leptonic emission model with the requirement of seed photons for inverse Compton scattering from both the BLR and the torus (\citealt{Paliya_2015}). Alternatively, during the epoch of the hard gamm-ray flare from 3C 279 in December 2013, the SED was fit by both lepto-hadronic model and two-zone leptonic model (\citealt{Paliya_2016}). This clearly indicates that even in the same source, different radiative processes contribute at different epochs, reflecting the complexities seen in emission from blazar jets. In another FSRQ, Ton 599, the $\gamma-ray$ emission site is found to be at the outer edge of the BLR (\citealt{Prince_2019}) based on leptonic model fit to the observed SED. Recently, the SED of the BL Lac Mrk 421 was found to be equally fit by all the models such as leptonic, hadronic and lepto-hadronic (\citealt{Cerruti_2020}). In the source Cen A, there are reports that the obsevations are fit by protron-synchrotron model (\citealt{Banik_2020}). Also, the recent detection of neutrions from the blazar TXS 0506+056 (\citealt{Aartsen_2018}) seems to favour lepto-hadronic over leptonic models (\citealt{Cerruti_2019}). It is believed that $\gamma-ray$ flares without counterparts in the optical band cannot be explained in the one-zone leptonic emission sceanrio and could favour hadronic models. Indeed, from analysis of a sample of FSRQs, \citet{Rajput_2019, Rajput_2020} have found that in a source there are epochs when there are optical flares without gamma-ray flares, gamma-ray flares without optical counterparts and correlated optical and gamma-ray flares, all of which are explainable in the leptonic scenario. Thus, recent observations on a handfuld of blazars clearly indicate that the exact reasons for the origin of high energy emission in blazar is complex, still not undestood and possibly future X-ray polarization observations could provide the needed observational constrants on the high energy emission process in blazars. Given the current scenario, it is of utmost importance to carry out timing and SED analysis of more and more number of blazars.

The blazar OQ 334 is classified as a flat spectrum quasar (FSRQ) in 4FGL (\citealt{2020ApJS..247...33A}) and it is at a redshift z = 0.6819 (\citealt{Hewett_2010}). It has been in the low $\gamma$-ray brightness state since 2008. After a decade it flared in the $\gamma$-ray band in 2018 (\citealt{Atel_12277}). Subsequently it was in a higher $\gamma$-ray brightness state and again flared in the $\gamma$-ray band in 2020 (\citealt{Atel_13382}). Observations were also available in the X-ray and optical/UV bands during the flaring epochs of the blazar from {\it Swift}. This source has not yet been studied for its $\gamma$-ray characteristics, however reports on its $\gamma$-ray flaring are available (\citealt{Atel_12277}, \citealt{Atel_12942}, \& \citealt{Atel_13382}). Therefore, in this work we carried out detailed timing and spectral analysis of the source during both its flaring periods as well as a quiescent period. \\
This paper is organized as follows. In Section 2, we describe the mulitwavelength data and reduction, the results are shown in Section 3, 4, \& 5, followed by the summary in the final Section.

\section{Multiwavelength Observations and Data Analysis}
The FSRQ OQ 334 was found to show two episodes of flaring, one in 2018 and the other in 2020, based on observations with the Large Area Telescope on board the {\it Fermi} Gamma-ray Space Telescope. To understand the nature of the source during its high activity states relative to its quiescent state necessitates creation of broadband SED, which in turn requires data at other wavelengths also. Towards this we looked into the $\gamma$-ray data from {\it Fermi-LAT}, X-ray from {\it Swift-XRT} and Ultraviolet and optical data from {\it Swift-UVOT}.  

\subsection{\textit{Fermi}-LAT}
\textit{Fermi-LAT} observes the galactic as well as the extragalactic sky in $\gamma$-ray. It is based on the pair conversion method, and the working energy
range is 20 MeV -- 500 GeV. It has a vast field of view (FoV) of about 2.4 sr (\citealt{Atwood_2009}), which scans 20\% of the sky at any time.
The total scanning period of the entire sky is around three hours. 
\textit{Fermi}-LAT is continuously monitoring the source OQ 334/B2 1420+32 since 2008.
LAT data in the energy range 100 MeV to 300 GeV were taken for the period December 2008 to February 2020. We followed the standard procedure for the reduction of $\gamma$-ray data as given in Science Tools\footnote{https://fermi.gsfc.nasa.gov/ssc/data/analysis/documentation/}. More details about the reduction can be found in \citet{Prince_2018}. We generated 3 day bin light curve covering a duration of 900 days from MJD 58000 to MJD 58900. This is shown in Figure 1. Each points in the light curve pertains to a test statistics (TS) greater than 9, which corresponds to a 3 $\sigma$ detection.
 
\begin{figure*}
 \includegraphics[scale=0.55]{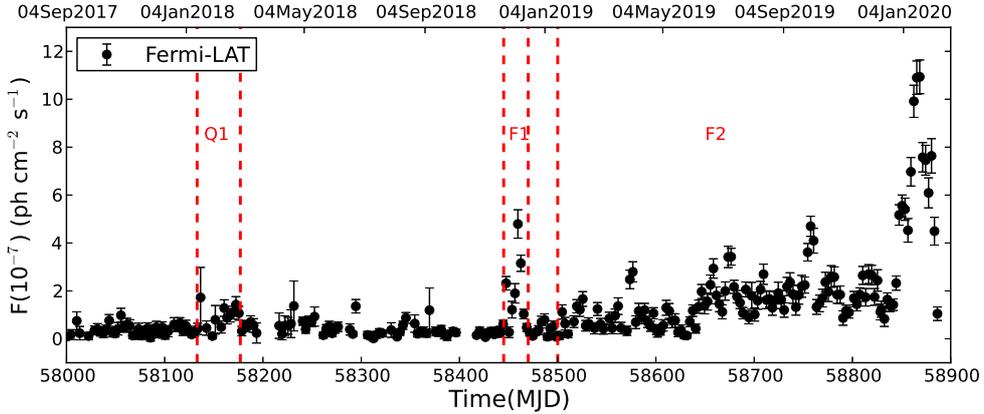}
 \caption{ The long-term $\gamma$-ray light curve of OQ 334. State Q1 represents the quiescent state, and F1 and F2 are the two flaring states.}
\end{figure*}

\begin{figure*}
\vspace{-15pt}
 \centering
 \includegraphics[scale=0.45]{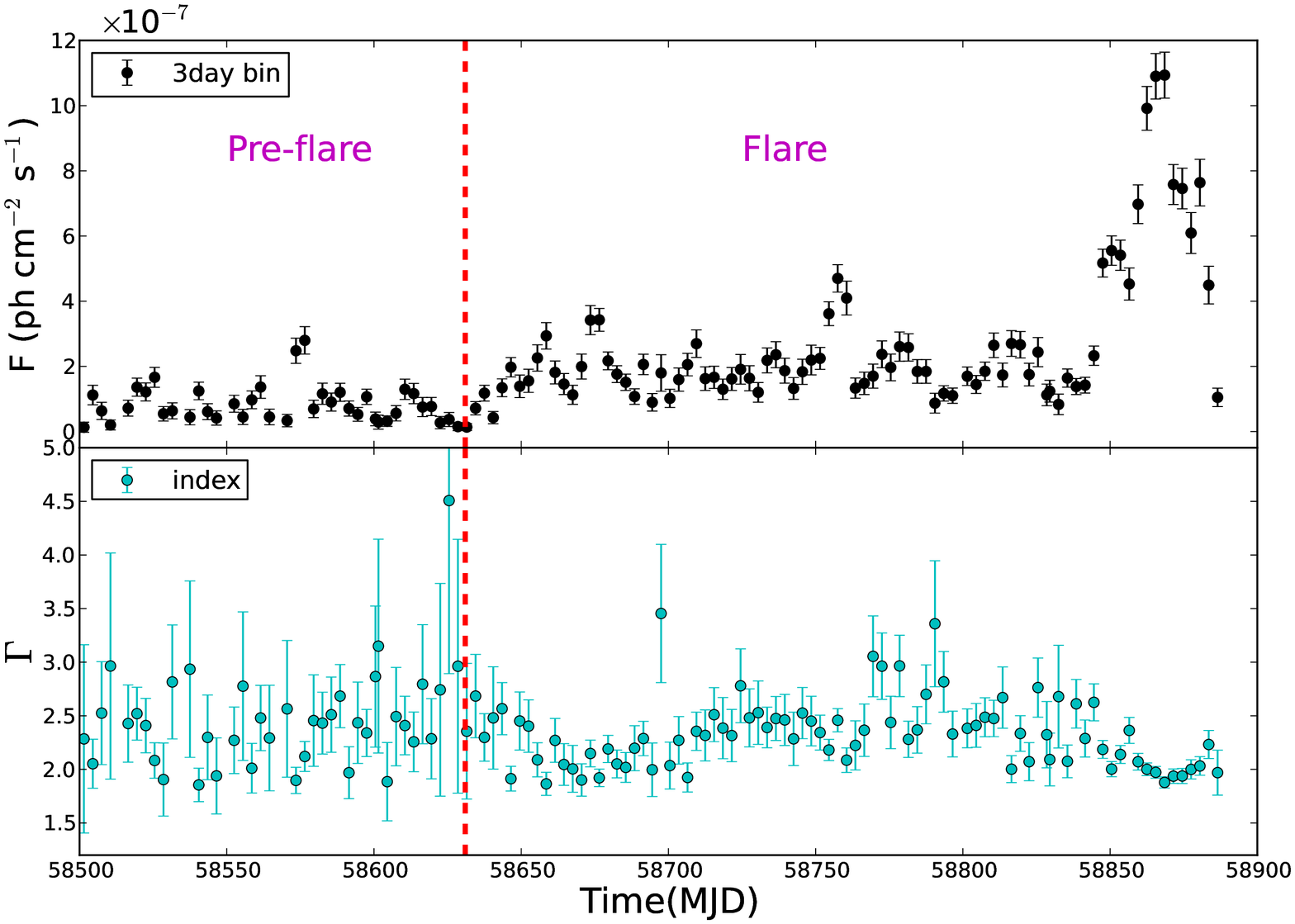}
 \includegraphics[scale=0.45]{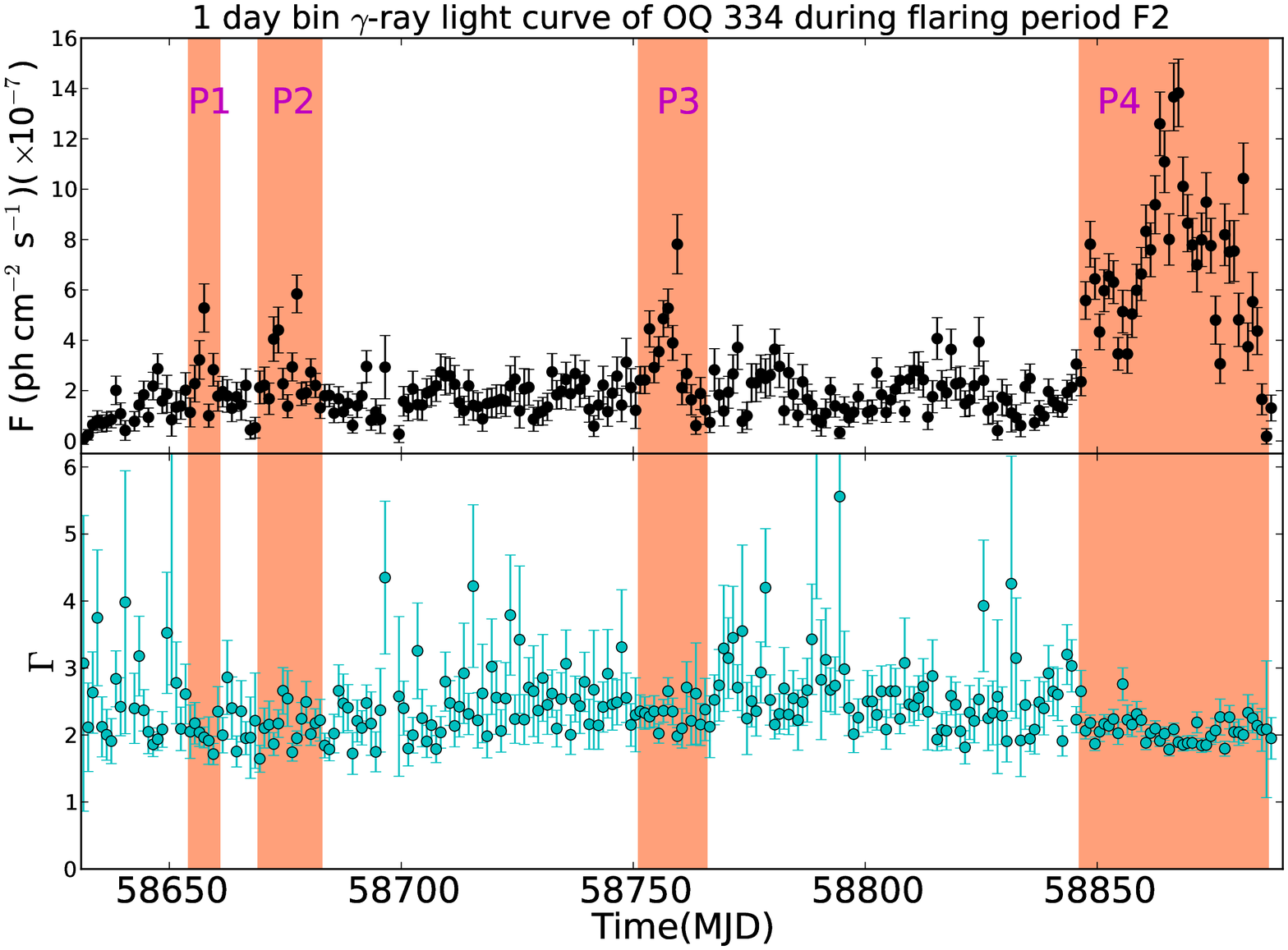}
 %\vspace{-50pt}
 \caption{The upper plot shows the $\gamma$-ray light curve of state F2 with the spectral index, divided into pre-flare and flare. The lower plot is the one-day binned $\gamma$-ray light curve of the flaring period. The color patches describe the various peaks observed in the flaring part.}
\end{figure*}

\subsection{Swift-XRT/UVOT}
\textbf{X-ray}-

The X-ray data used in this work is from {\it Swift/XRT} in the energy range of 0.3 - 10 keV. X-ray data from {\it Swift} was not available for most of the duration of the $\gamma$-ray light curve, however sparsely available during the two flaring periods of OQ 334. The log of the X-ray observations used in this work is given in Table 1. For all the observations given in Table 1, we used the task xrtpipeline to generate clean event files. We used
the CALDB version 20160609 for {\it Swift-XRT} analysis. The clean event files were further processed using xselect in XSPEC (\citealt{1996ASPC..101...17A}). The events from the source were selected from a circular region of radius 12 arcsec. The background was selected from a region of similar size but away from the source. We modelled the generated spectra using the simple power law model F(E) $\propto$ E$^{\Gamma}$. For spectral fit we binned the data to have a minimum of 20 counts/bin. For the power law fit, we used a fixed galactic absorption column density $n_H$ = 1.10 $\times$ 10$^{20}$ cm$^{-2}$ taken from \citet{Kalberla_2005}. We generated the spectrum file for each observations and if in a particular flaring period we have more than one observations, we combined the spectra in addspec\footnote{https://heasarc.gsfc.nasa.gov/ftools/caldb/help/addspec.txt}. In addspec, we added the source spectra along with the redistribution matrix files (RMF) and the ancillary response files (ARF) for different observations. Similarly, the background spectra
from different observations were added in mathpha\footnote{https://heasarc.gsfc.nasa.gov/ftools/caldb/help/mathpha.txt}. The combined X-ray spectrum from each period (quiescent-sate, flare-1, and P4) was finally used in the multi-wavelength SED modeling.\\
%To get the spectra for one particular period for SED analysis, we generated the combined spectra using addspec in XSPEC.\\
\\
\textbf{UV and Optical}-

In the optical and ultra-violet (UV) bands, we used data from the {\it Swift-UVOT} (\citealt{Roming_2005}) in the filters U, B, V, W1 and W2. For each of these filtes, we added the observations accumulated over a given period using the task UVOTIMSUM and we derived the magnitude of the target blazar using UVOTSOURCE. We corrected the magnitudes for galactic extinction following \citet{Schlafly_2011} and converted the extinction corrected magnitudes to fluxes using the zero points (\citealt{Breeveld_2011}) and conversion factors given in \citet{Larionov_2016}.

\begin{table}
\centering
\caption{Log of the {\it Swift} observations during all the states (Q1, F1, and P4).}
 \begin{tabular}{ccc p{1cm}}
 \hline
 && \\
 Observatory & Obs-ID & Exposure (ks)\\
 && \\
 \hline
 Q1 \\
 \hline
 Swift-XRT/UVOT & 00010520001 & 1.0\\
 Swift-XRT/UVOT & 00010520002 & 2.9\\
 \hline
 F1\\
 \hline
 Swift-XRT/UVOT & 00010520004 & 1.8\\
 Swift-XRT/UVOT & 00010520005 & 2.1\\
 Swift-XRT/UVOT & 00010520006 & 1.9\\
 \hline
 P4 \\
 \hline
 Swift-XRT/UVOT & 00010520014 & 2.5\\
 Swift-XRT/UVOT & 00010520015 & 1.6\\
 Swift-XRT/UVOT & 00010520016 & 1.9\\
 Swift-XRT/UVOT & 00010520017 & 2.0\\
 Swift-XRT/UVOT & 00010520018 & 1.6\\
 Swift-XRT/UVOT & 00010520019 & 1.5\\
 Swift-XRT/UVOT & 00010520022 & 1.8\\
 Swift-XRT/UVOT & 00010520024 & 1.8\\
 Swift-XRT/UVOT & 00010520025 & 1.8\\
 Swift-XRT/UVOT & 00010520026 & 2.0\\
 Swift-XRT/UVOT & 00010520027 & 1.2\\
 Swift-XRT/UVOT & 00010520028 & 1.7\\
 Swift-XRT/UVOT & 00010520029 & 2.0\\
 Swift-XRT/UVOT & 00010520030 & 2.4\\
 \hline
 
 \end{tabular}
 \label{Table:T1}
\end{table}  

\subsection{Radio data at 15 GHz}
In the radio band we used the data at 15 GHz from the Owens Valley Radio Observatory (OVRO; \citealt{Richards_2011}). The blazar OQ 334 is part of the OVRO monitoring program and therefore data at 15 GHz with a time resolution of about 2 weeks is available for most of the duration of the $\gamma$-ray light curves.

\section{Results: Temporal and Spectral analysis}
A detailed temporal and spectral study has been done using the multi-wavelength data from the \textit{Fermi}-LAT, and the Swift-XRT/UVOT telescope. The archival data from OVRO is used to perform the correlation study with the $\gamma$-ray to examine the possible location of their emission regions.

\begin{table}
 \centering
 \caption{Details about the various states recongnized in this study.}
 \begin{tabular}{cccc}
  \hline
  States & MJD start & MJD end & Duration (days) \\
  \hline
  Q1    & 58133 & 58177 & 44 \\
  F1    & 58445 & 58470 & 25 \\
  F2/pre-flare & 58500 & 58631 &  131 \\
  F2/ flare & 58631 & 58887 & 256 \\
  P1 & 58654 & 58661 & 7 \\
  P2 & 58669 & 58683 & 14 \\
  P3 & 58751 & 58766 & 15 \\
  P4 & 58846 & 58887 & 41 \\
  \hline
  
 \end{tabular}

\end{table}

\subsection{$\gamma$-ray flux variability}
We show in Figure 1 the three day binned $\gamma$-ray light curve of OQ 334. As can be seen, the source was in a low $\gamma$-ray brightness state till 2018, when it first showed a $\gamma$-ray flare with a three day binned $\gamma$-ray flux of 4.79$\pm$0.59 ($\times$10$^{-7}$ ph cm$^{-2}$ s$^{-1}$) and an average $\gamma$-ray photon index of 1.96$\pm$0.10, this has also been reported in \citet{Atel_12277} and shown in Figure 1 as F1 (Flare-1). The source remained in the low brightness state for a few days after the flare and then it showed a steady increase in the $\gamma$-ray brightness level. Superimposed on the high $\gamma$-ray
brightness level, we noticed several flares that also includes a major flare in early
2020. From the three day binned light curve, we identified three major regions, a quiescent state Q1, a flaring state F1 and a higher brightness state F2. The higher brightness state was further divided into pre-flare (MJD 58500 - 58631) and flaring period (MJD 58631 - 58887), based on the average flux seen in the 3 day binned $\gamma$-ray light curve (Figure 2). For pre-flare and flare periods we found average flux values of 6.09$\pm$0.36 ($\times$10$^{-8}$ and 1.72$\pm$0.03 ($\times$10$^{-7}$ ph cm$^{-2}$ s$^{-1}$) respectively. To substantiate this division based on the mean flux values, we also calculated the fractional flux variability (F$_{Var}$; \citealt{Prince_2018}). We found F$_{Var}$ values of 0.61$\pm$0.05 and 0.81$\pm$0.02, for the pre-flare and flare periods respectively.

To further characterise the flux variability pattern in the flaring period, we generated 1 day binned $\gamma$-ray light curve. From this 1 day binned light curve, we identified four high flux states (with flux exceeding $\sim$ 4$\times$10$^{-7}$ ph cm$^{-2}$ s$^{-1}$ ) denoted as P1, P2, P3 and P4. The colour patches in Figure 2 denote the total duration of each of these flux states. The details about all the states and their periods are mentioned in Table 2.
Further, we also generated 12 hour binned $\gamma$-ray light curve to model the variations in the high flux states with a sum of exponential functions. This function is used to estimate the rising and decaying time of the peaks observed within the high state periods. The functional form of the sum of exponentials is given below,
\begin{equation}
 F(t) = 2F_0 \left[ \exp \left( \frac{t_0 - t}{T_r} \right) + \exp \left( \frac{t - t_0}{T_d} \right) \right]^{-1}
\end{equation} 
where T$_r$ and T$_d$ are the rise and decay times of the peaks, respectively, and the peak amplitude is approximated as F$_0$ measured at time t$_0$. The fits to the lightcurve for all the high states (P1, P2, P3, and P4) are shown in Figure 3, and the corresponding fitted parameters are given in Table 3.
Various peaks observed in the high states P1, P2, P3, and P4 in Figure 3 from left to right are denoted in the serial number (1,2,3,..etc.) in Table 3. The temporal fitting for the high states P1, P2, and P3 was done for the 12 hr binned light curve. However, in the case of high state P4, the 12 hr binned light curve has large error bars, and hence we considered the one-day binned light curve for temporal study. For the first three cases, we also considered the constant flux state (shown in grey) while doing the temporal fitting, which shows the flux level before and after the peaks. 

\subsubsection{Spectral variations}
A harder when brighter trend is generally seen in the high energy $\gamma$-ray emission from blazars (Ton 599; \citealt{Prince_2019}, 3C 279; \citealt{Prince_2020}). To investigate spectral variations if any, we show in Figure 4, the variation of $\Gamma$ with the observed $\gamma$-ray brightness during the states P1, P2, P3 and P4. We did not find any significant spectral variations with brightness.

\begin{table}
 \centering
\caption{The rise and decay times estimated from Equation (1) for all the peaks. The peak flux F$_0$ is in units of ($\times$10$^{-7}$) ph cm$^{-2}$ s$^{-1}$.}
 \begin{tabular}{ccccc p{0.1cm}}
 \hline \
 Peaks & t$_0$ & F$_0$ & T$_r$ & T$_d$ \\
    &  (MJD) & & (hr) & (hr) \\
\hline \
P1 & & & & \\
1 & 58655.25 & 3.12$\pm$1.21 & 13.93$\pm$2.99 & 6.22$\pm$1.76 \\
2 & 58657.25 & 7.34$\pm$1.43 & 13.67$\pm$1.47 & 5.56$\pm$0.78 \\
3 & 58659.25 & 3.12$\pm$1.03 & 14.90$\pm$3.35 & 17.69$\pm$5.20 \\
\hline \
P2 & & & &\\
1 & 58673.25 & 5.46$\pm$1.26 & 14.51$\pm$3.43 & 8.48$\pm$2.75 \\
2 & 58677.25 & 7.59$\pm$1.15 & 8.13$\pm$1.46 & 7.76$\pm$1.40 \\
3 & 58680.75 & 3.03$\pm$0.82 & 7.05$\pm$2.96 & 10.77$\pm$4.05 \\
\hline \
P3 & & & &\\
1 & 58753.75 & 6.07$\pm$1.23 & 6.73$\pm$1.92 & 8.51$\pm$2.93 \\
2 & 58755.25 & 4.20$\pm$0.88 & 4.70$\pm$4.34 & 4.85$\pm$4.32 \\
3 & 58757.25 & 6.65$\pm$1.19 & 15.03$\pm$4.07 & 11.78$\pm$3.58 \\
4 & 58759.25 & 9.29$\pm$1.63 & 6.72$\pm$2.01 & 9.34$\pm$2.12 \\
\hline  \
P4 & & & & \\
1 & 58848.50 & 7.81$\pm$0.90 & 29.12$\pm$5.03 & 26.49$\pm$11.93 \\
2 & 58852.50 & 6.54$\pm$0.89 & 23.66$\pm$15.68 & 27.33$\pm$9.80 \\
3 & 58863.50 & 12.60$\pm$1.26 & 125.17$\pm$15.19 & 10.69$\pm$4.67 \\
4 & 58867.50 & 13.82$\pm$1.34 & 18.87$\pm$5.01 & 90.34$\pm$14.66 \\
5 & 58873.50 & 9.49$\pm$1.17 & 27.75$\pm$12.08 & 22.65$\pm$9.50 \\
6 & 58877.50 & 8.19$\pm$1.23 & 15.15$\pm$7.67 & 88.66$\pm$10.45 \\

\hline \\
 \end{tabular}
\end{table}

\begin{figure*}
 \includegraphics[scale=0.35]{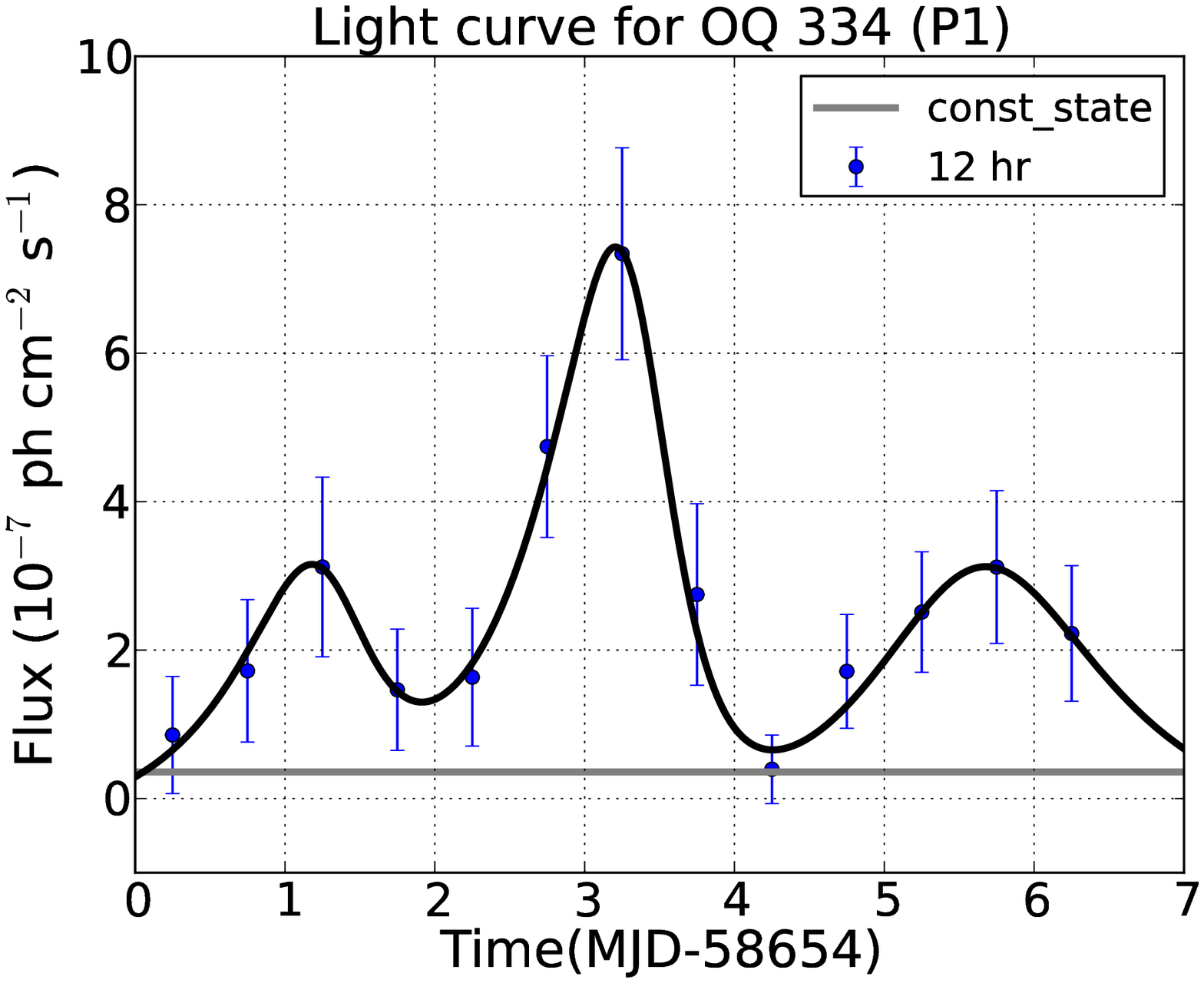}
 \includegraphics[scale=0.35]{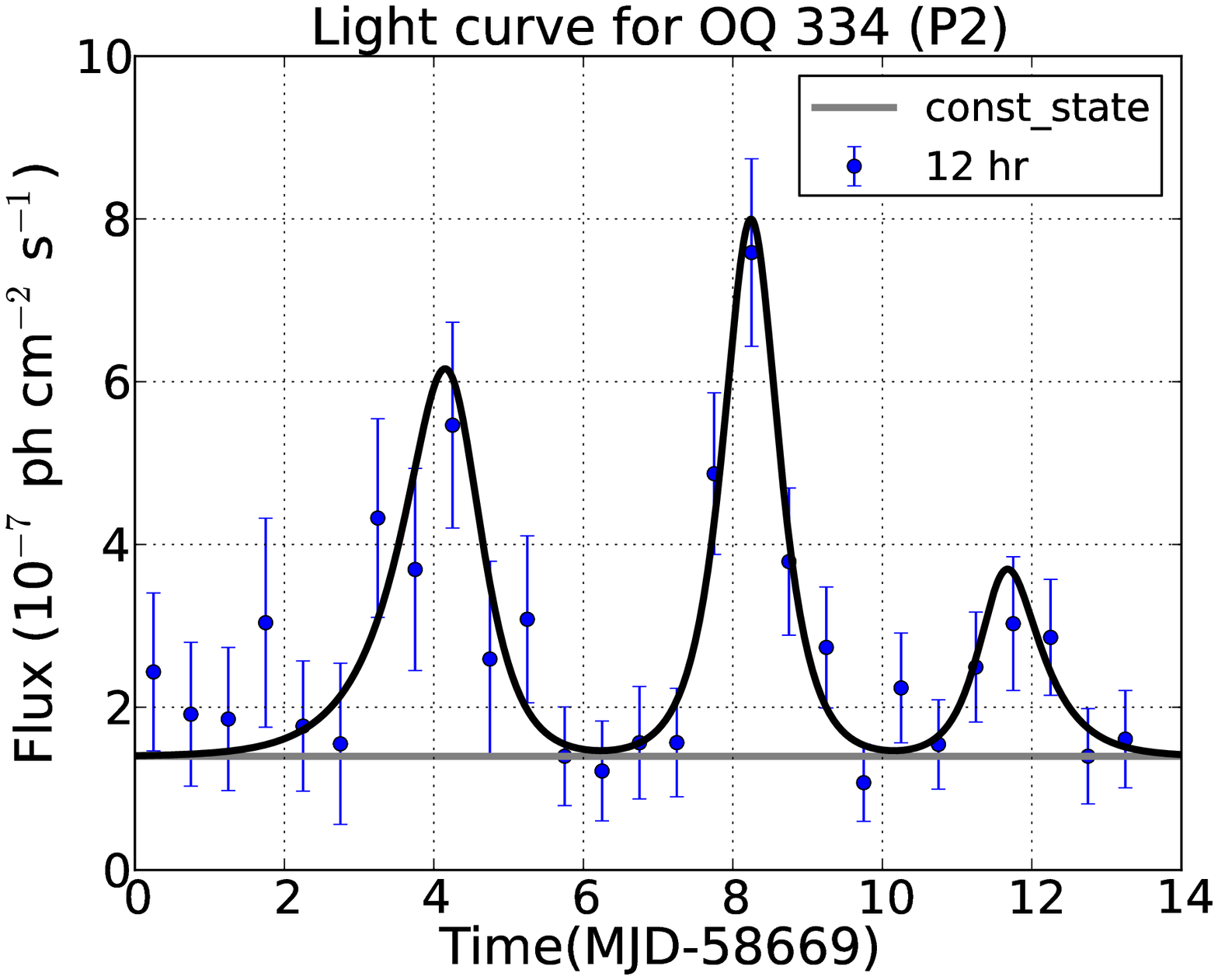}
 \includegraphics[scale=0.35]{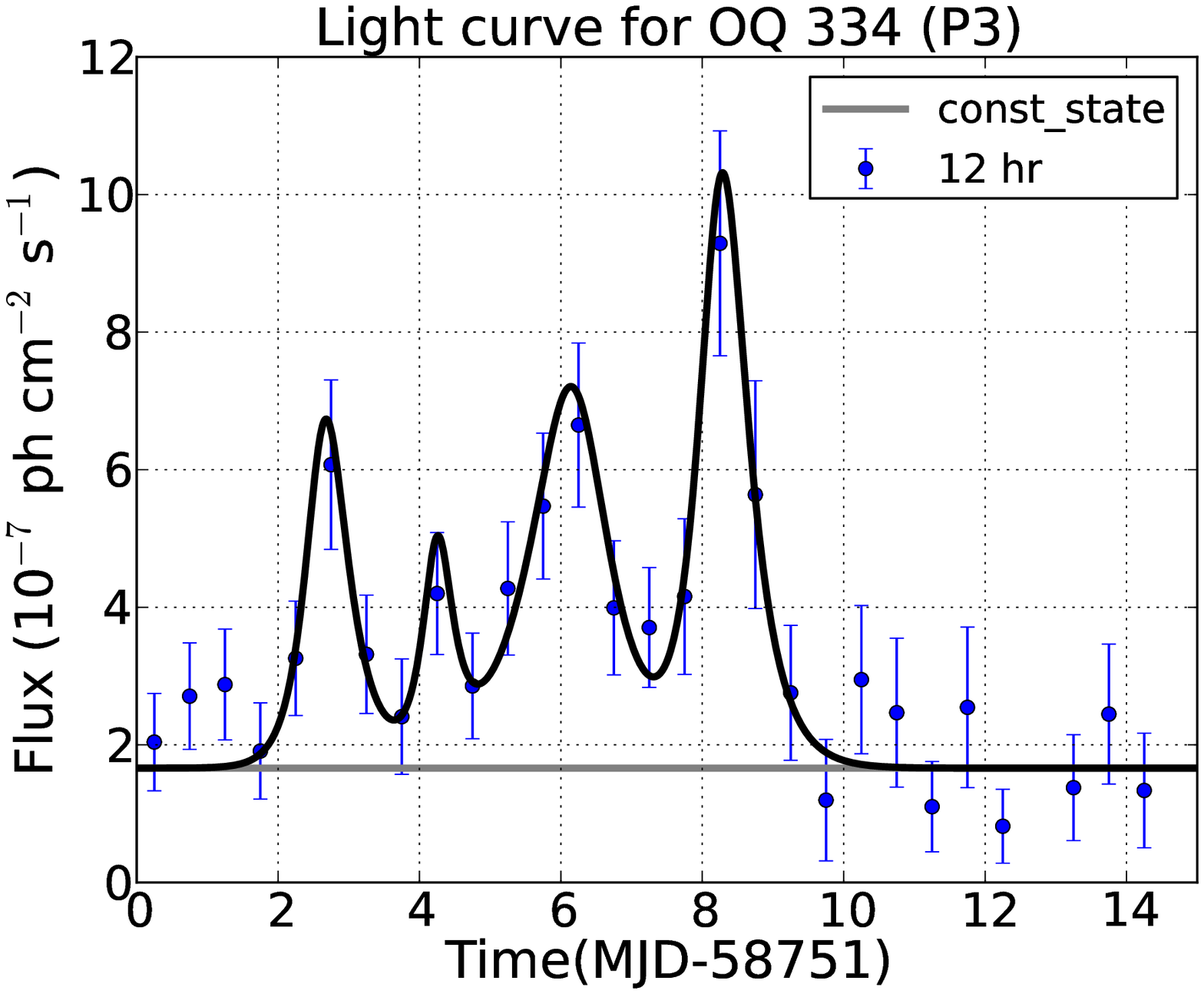}
 \includegraphics[scale=0.35]{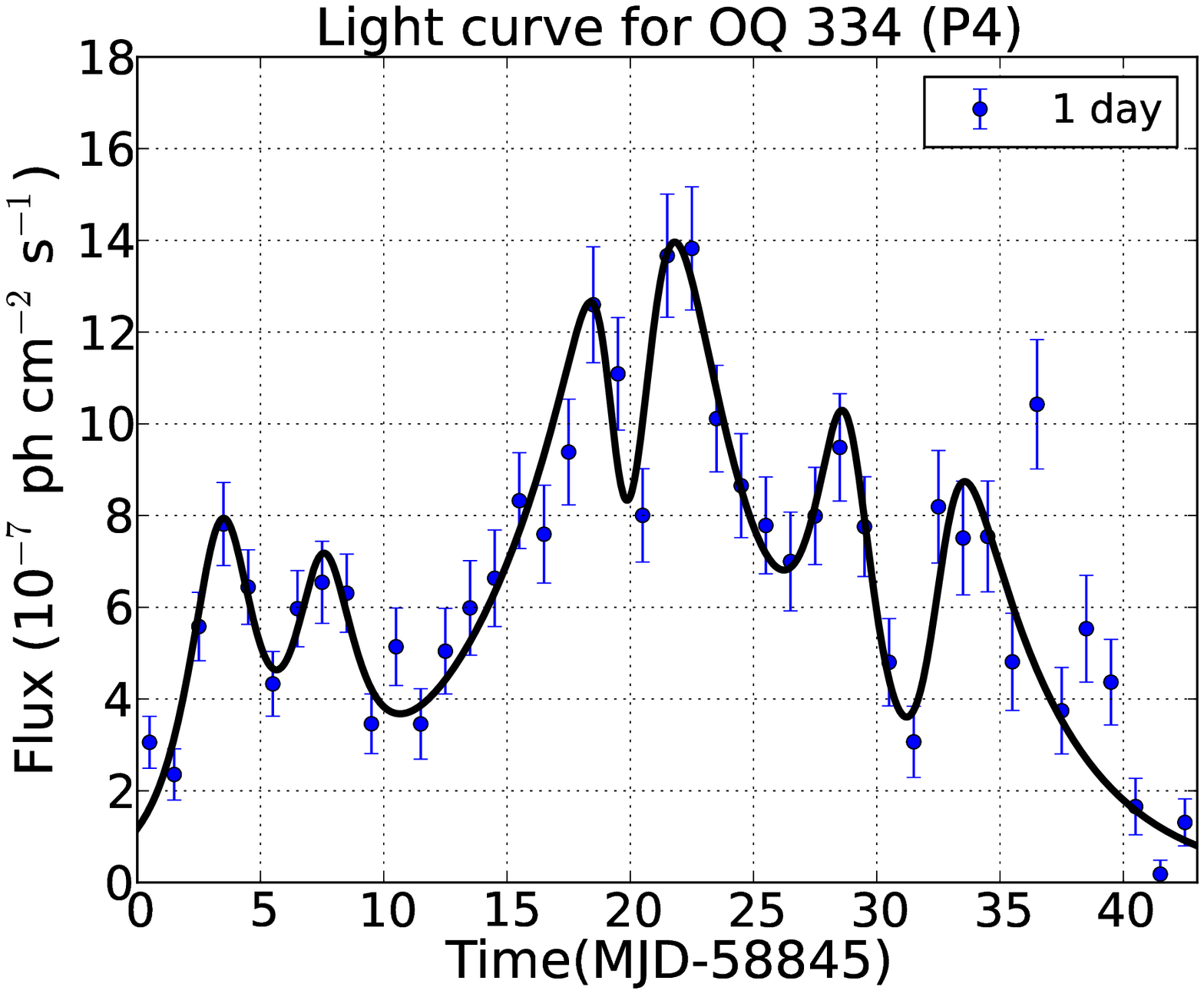}
 \caption{Light curve fitting of different flares observed in OQ 334. }
\end{figure*}

\begin{figure}
 \centering
 \includegraphics[scale=0.55]{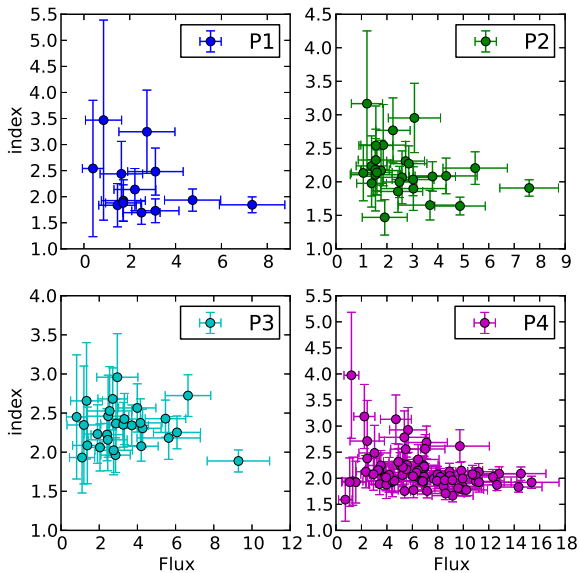}
 \caption{The $\gamma$-ray photon spectral index of all high states from 12 hr bin light curve, with respect to the observed flux. The brighter-when-harder behavior is not much clear in this case unlike most of the FSRQ type blazar .} %The x-axis is in units of ph cm$^{-2}$ s$^{-1}$.}
\end{figure}

\subsection{Highest energy photon}

In the FSRQ category of AGN, detection of $\gamma$-ray photons with energy > 20 GeV ( \citealt{Liu_2006}) suggests the location of the $\gamma$-ray emission region outside the broad line region (BLR). In such instances, the high energy photons, in the leptonic scenario are the result of inverse Compton scattering of photons from the dusty torus by the relativistic electrons in the jet. We estimated the number of high energy photons during the higher brightness state of the source and this is shown in Figure 5. We found two instances when $\gamma$-ray photons of energy > 70 GeV with the probability of being from the source is greater than 99\%. By comparing Figure 5 with the light curve shown in Figure 2, we conclude that the photon with energy $\sim$ 72 GeV was detected just before the high state P1, while the photon of energy $\sim$ 77 GeV was detected during P4. Detection of such high energy photons point to the location of the $\gamma$-ray emission site just at the boundary of BLR or the inner edge of the torus during P4.

\begin{figure*}
 \centering
 \includegraphics[scale=0.48]{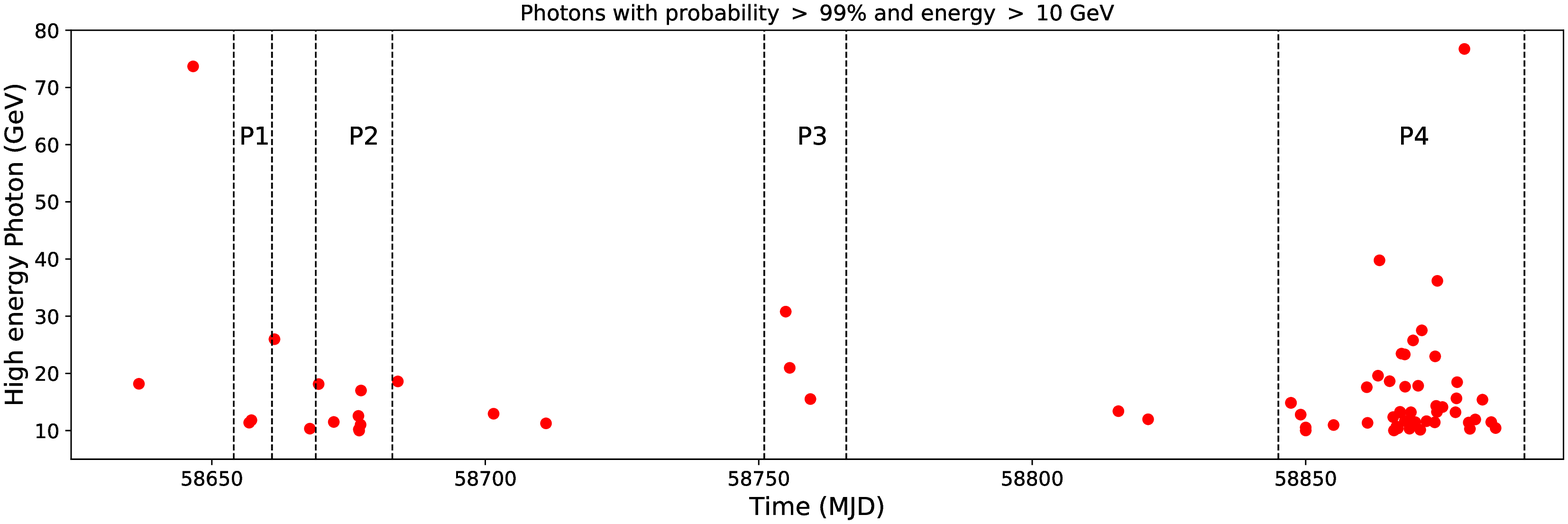}
 \caption{The arrival of high energy photon with probability $>$ 99$\%$ for being from the source are shown here. We have considered photons of
 energy greater than 10 GeV.}
\end{figure*}
 
\begin{figure*}
\centering
 \includegraphics[scale=0.283]{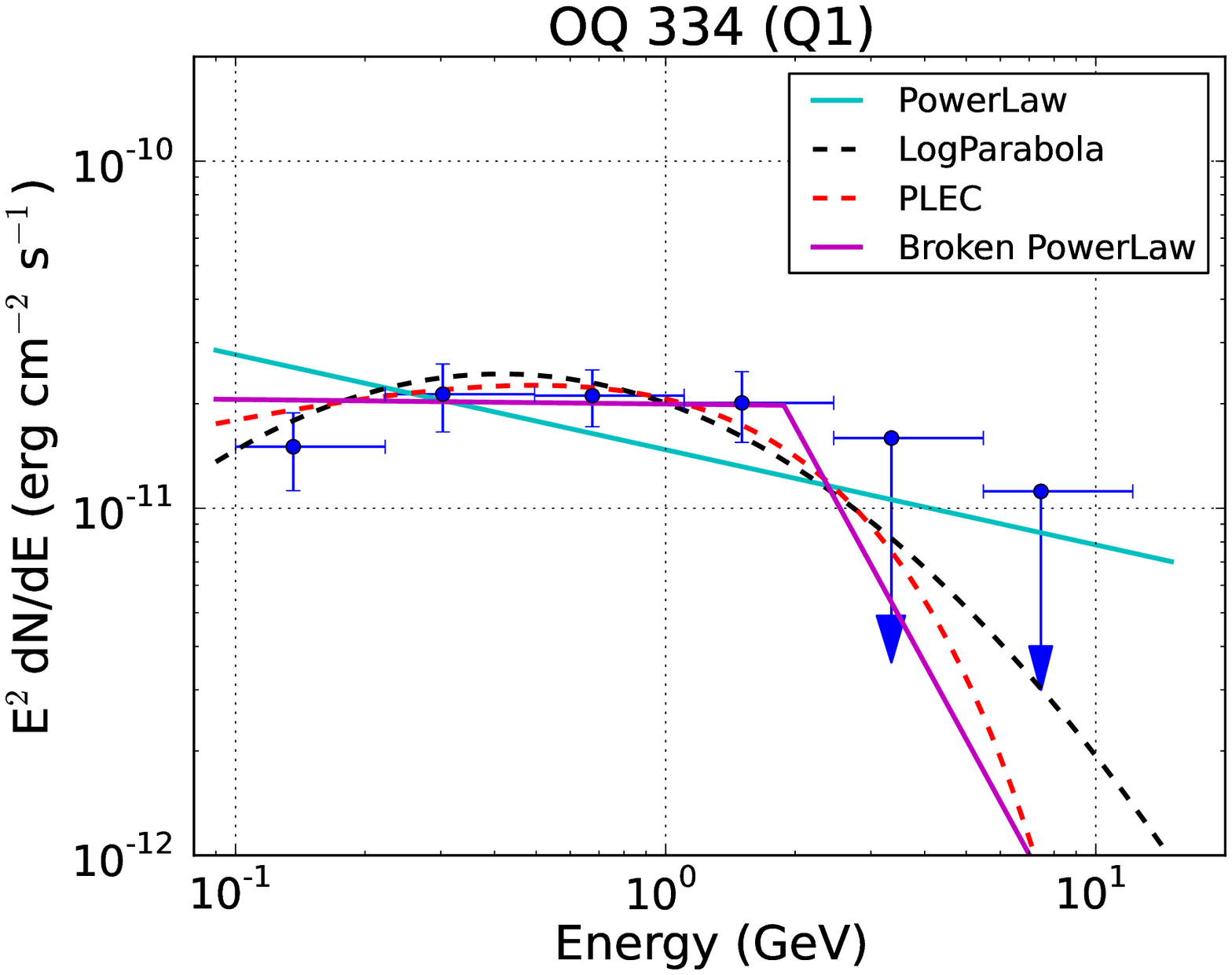}
 \includegraphics[scale=0.283]{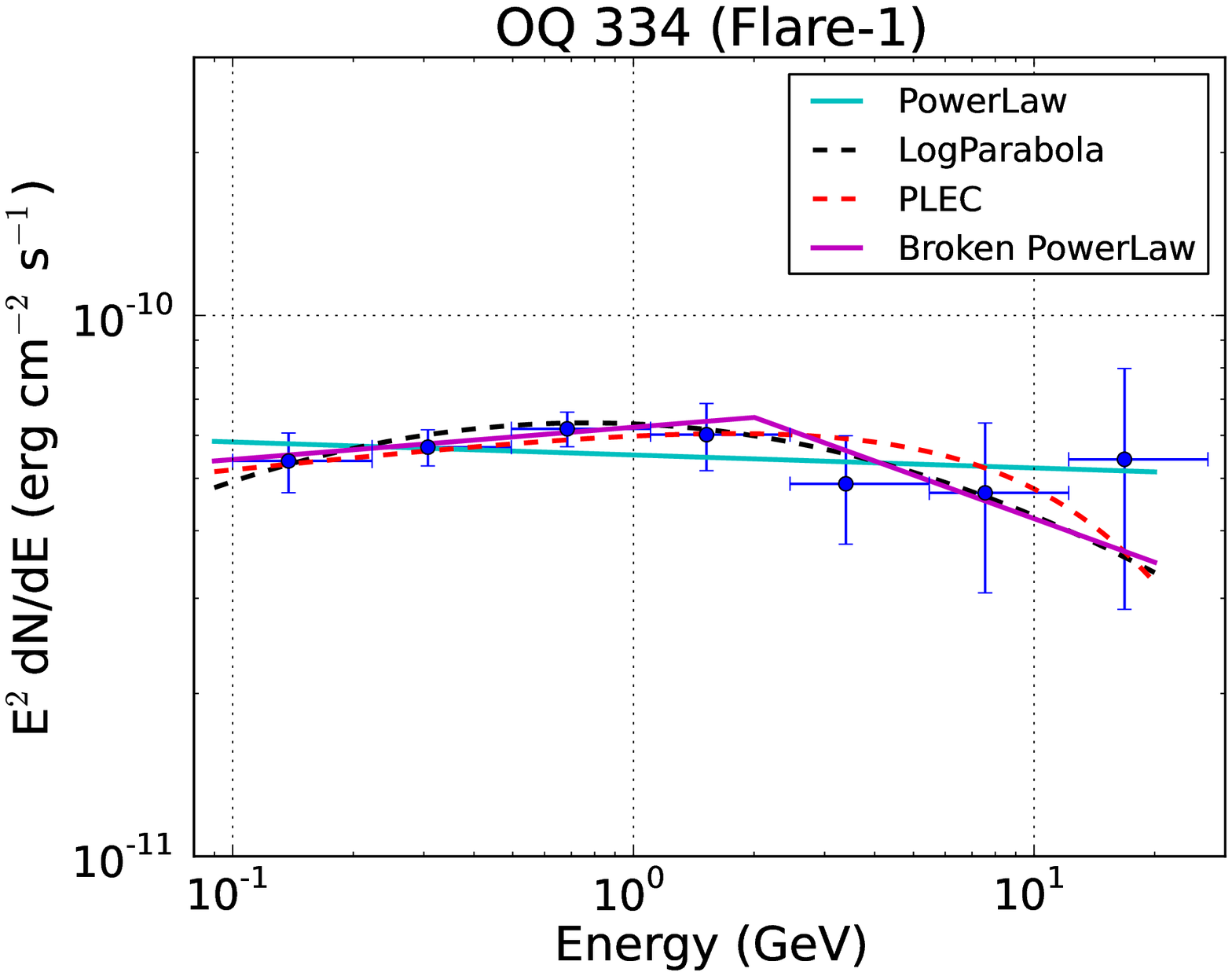}
 \includegraphics[scale=0.283]{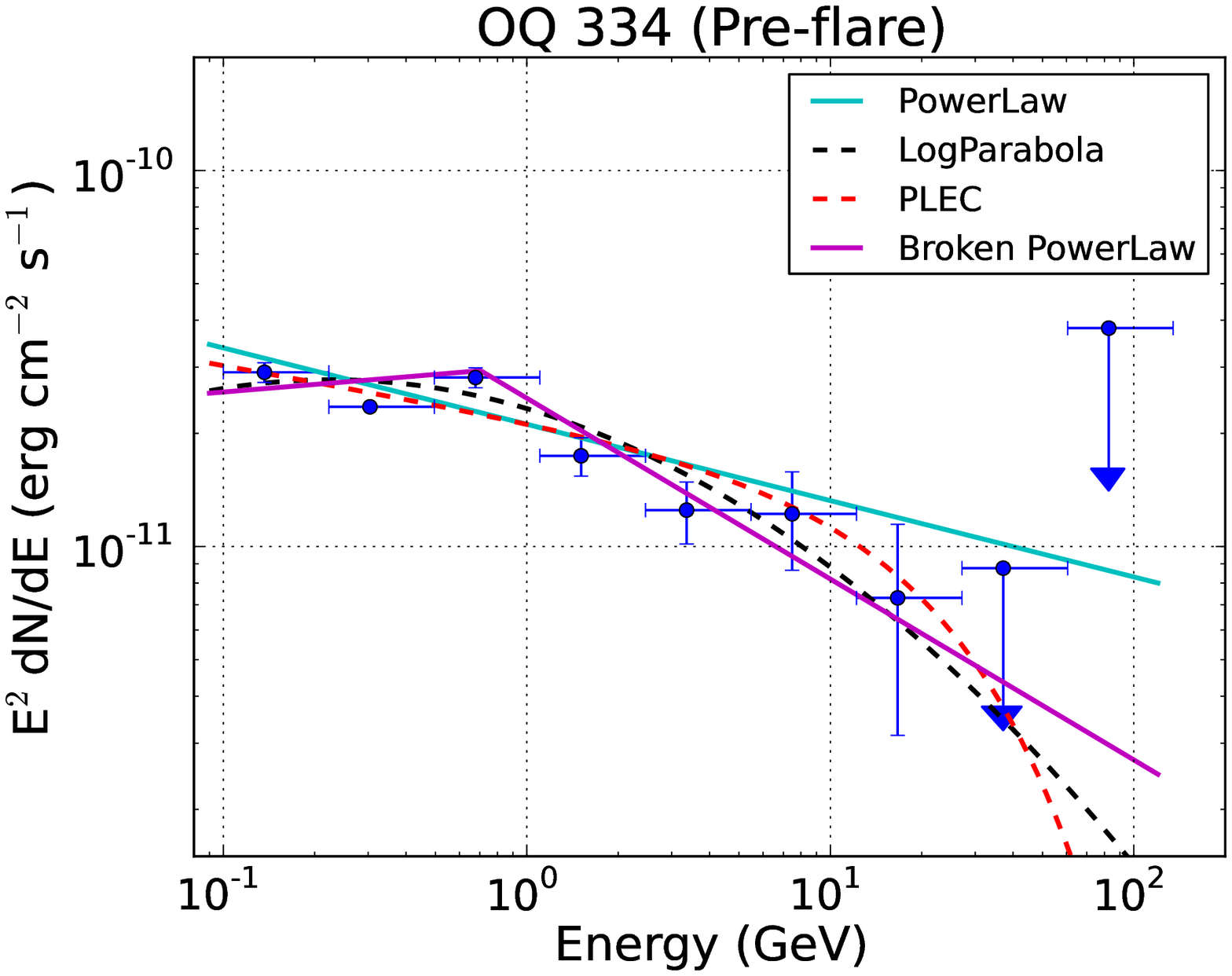}
 \includegraphics[scale=0.283]{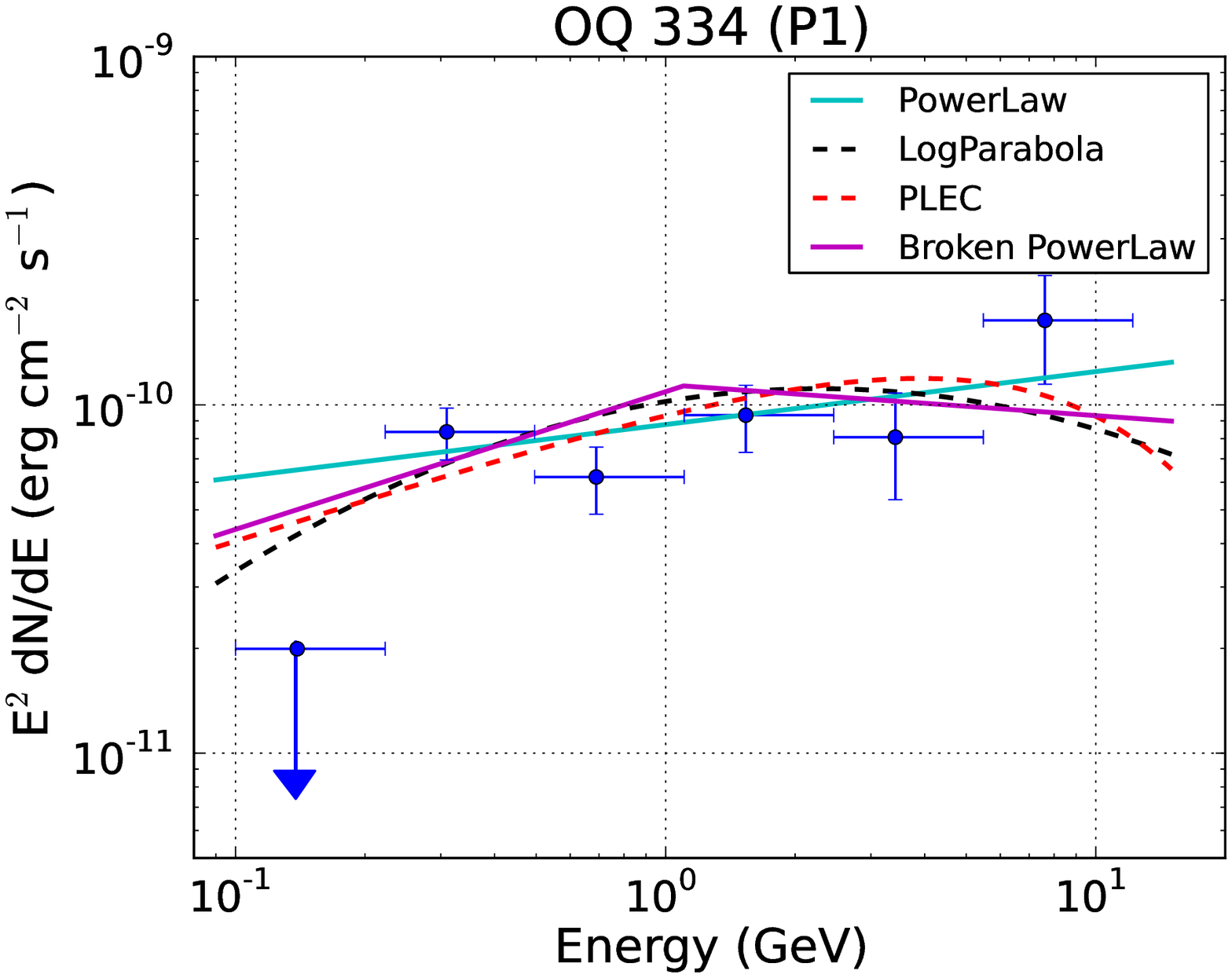}
 \includegraphics[scale=0.283]{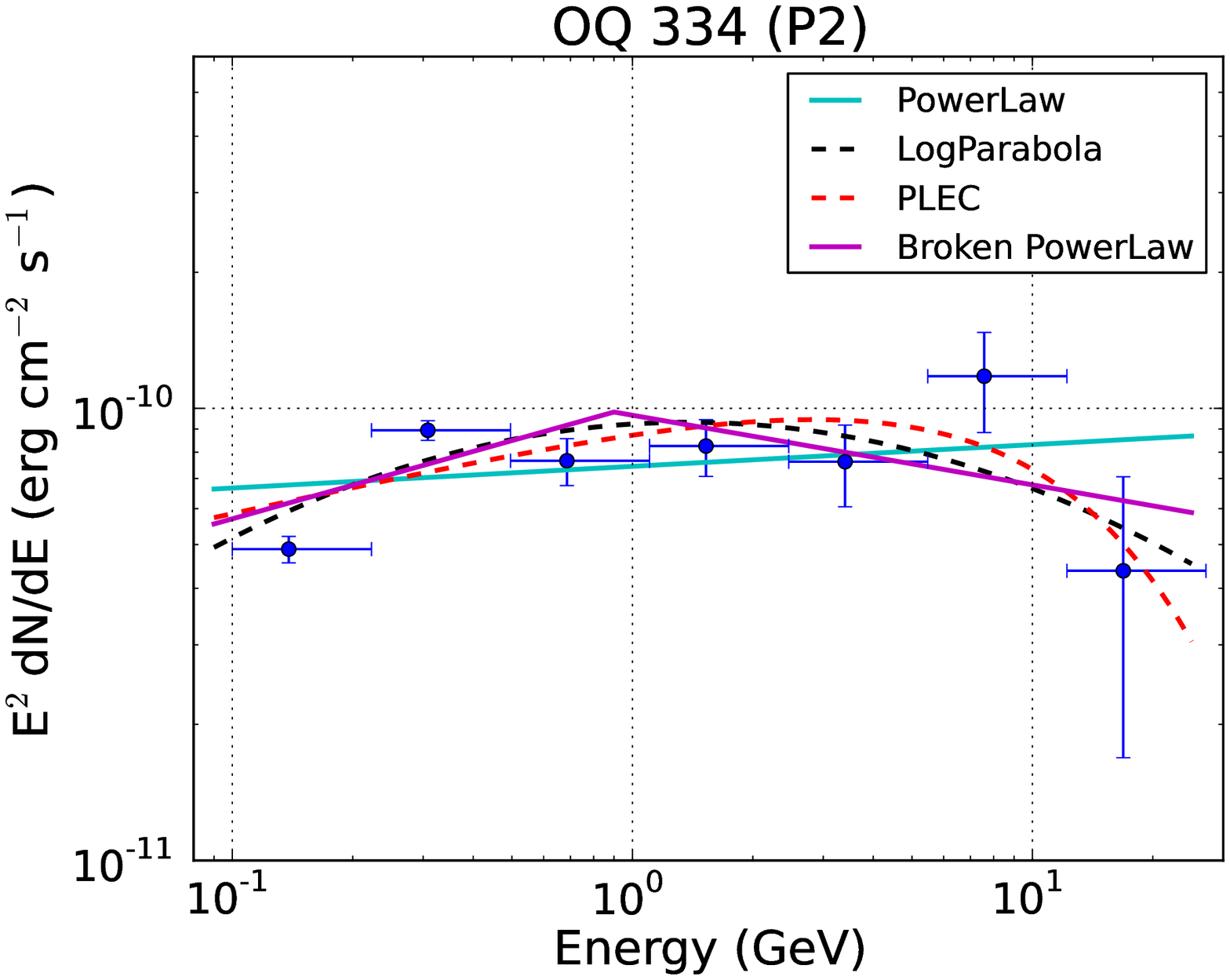}
 \includegraphics[scale=0.283]{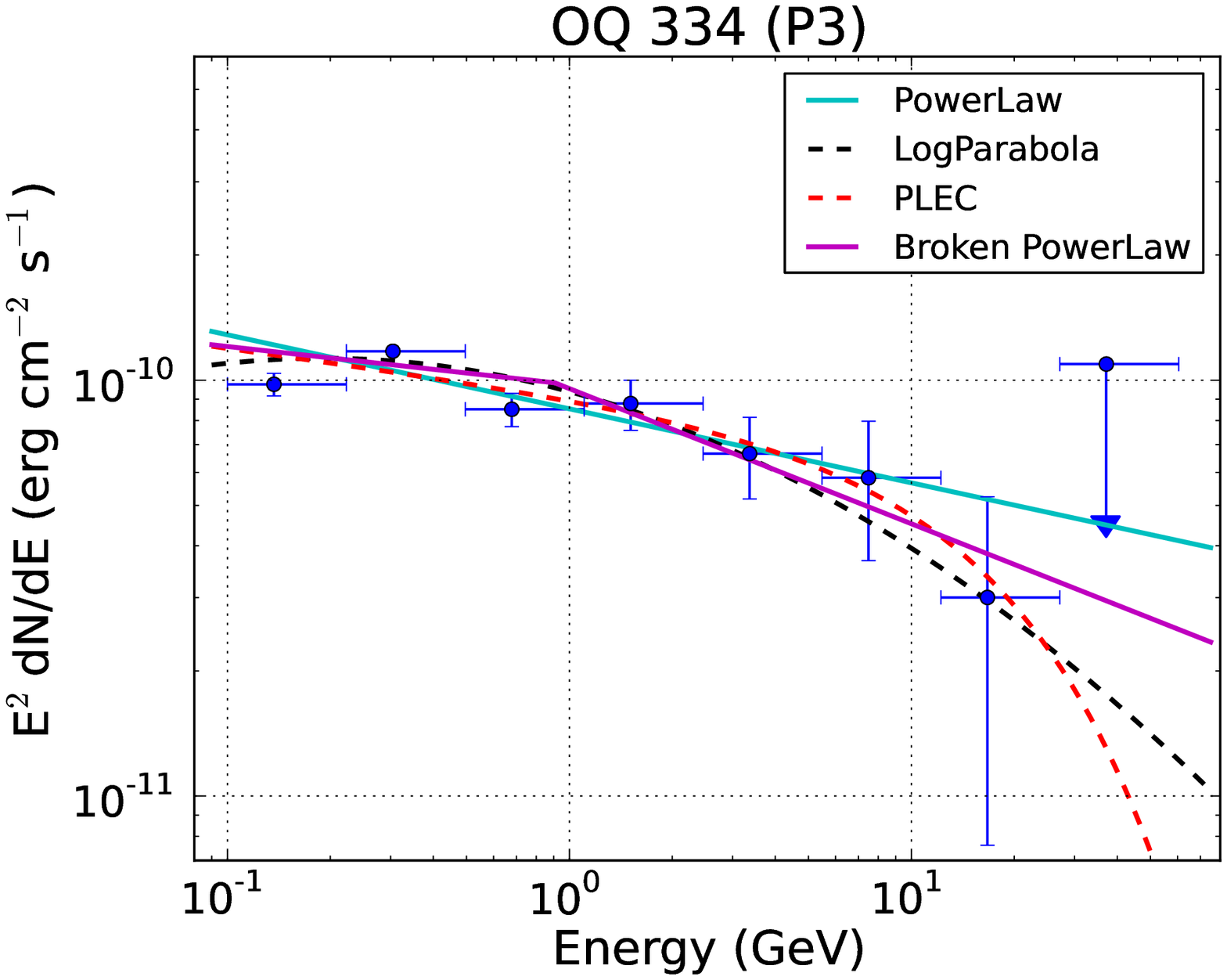}
 \includegraphics[scale=0.283]{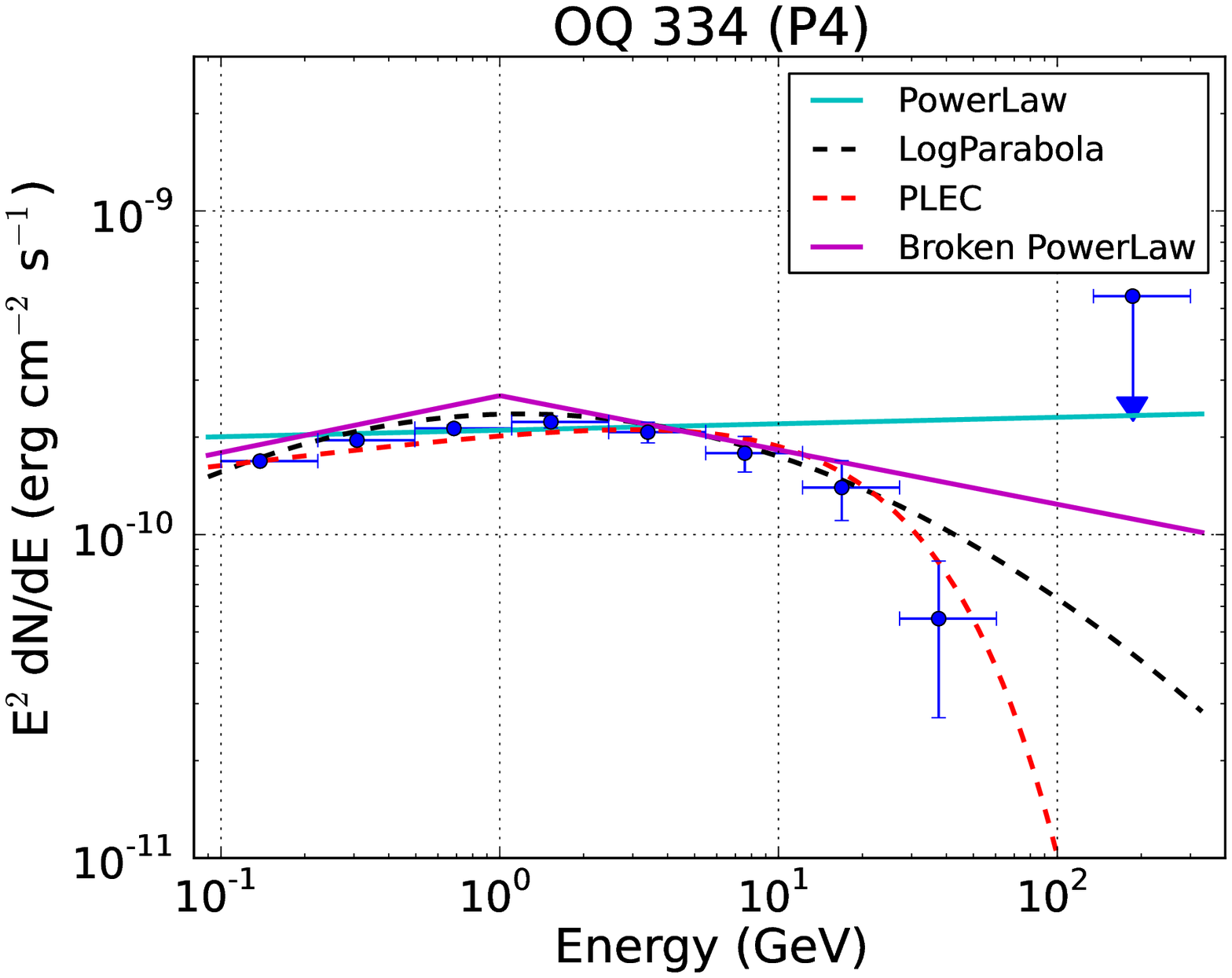}
 \caption{$\gamma$-ray spectral energy distributions (SED) for all the four flaring states and one long pre-flare state along with state Q1 and F1. Four different spectral
 models are used to fit the spectrum.}
 \end{figure*}
 
\subsection{$\gamma$-ray Spectra}
We generated the $\gamma$-ray spectrum for all the states identified in Figure 1 and Figure 2. The spectra were created using \textit{likeSED.py} a python code provided by {\it Fermi} Science Tools. We carried out likelihood analysis on the spectral data points using power law (PL), log parabola (LP), broken power law (BPL) and power law with exponential cut-off (PELC) models. The $\gamma$-ray spectra along with the model fits are shown in Figure 6 and the corresponding best fit model parameters are given in Table 4. We calculated TS$_{curve}$ = 2(log L(LP/BPL/PLEC) - log L(PL)), where L represents the likelihood function \citep{Nolan_2012}, to arrive at the suitability of the LP, BPL, and PLEC models over the PL model to describe the data. The best spectral model favours a value larger than TS$_{curve}$ = 16 which is significant at the 4 sigma level (\citealt{Mattox_1996}). 

The TS$_{curve}$ reveals presence of curvature or break in the spectrum, and which could be caused by the absorption of high energy photons ($>$ 20 GeV; \citealt{Liu_2006}) by the BLR assuming the emitting region is located within the BLR. However, if the emitting region is located outside the BLR a nice power law spectral behavior is expected. 

The various models and their corresponding parameters are shown in Table 4. 
The states, Q1, F1, pre-flare, P1, P2, P3 and P4 prefer models deviant from PL behaviour. 
The states where the PLEC best fits the spectra appears to have similar cut-off energy between 20-30 GeV, which is also the cut-off put by $\gamma$-$\gamma$ absorption (\citealt{Liu_2006}) within the BLR, and hence it is likely that the $\gamma$-ray emission region associated with these states is present at the outer boundary of the BLR. However, a recent study of 106 FSRQs by \citet{Costamante_2018} suggests that the smooth cut-off seen in the $\gamma$-ray spectrum between a few GeV to a few tens GeV is most probably the result of the end of the accelerated particle distribution.

The PL photon spectral index suggests a significant spectral hardening as the source transits from pre-flare to flare state (P1, P2, P3, \& P4) and it is also true when the source changes from state Q1 to F1. The cut-off energy (E$_{cutoff}$) in the PLEC model is different for different flaring states and this is compatible with the detection of high energy photons in various states (Figure 5). So, E$_{cutoff}$ for P1, P2, P3, \& P4 are $\sim$10 GeV, $\sim$20 GeV, $\sim$24 GeV, \& $\sim$28 GeV, and comparing these values to that in Figure 5 suggests that no high photons greater than these energies are detected in all the various states except P4. A high energy photon of energy greater than 70 GeV is found during P4, which suggests that this photon might be produced at the boundary or beyond the BLR. The BPL fit to the spectrum reveals that the break energy (E$_{break}$) is almost constant (at $\sim$1 GeV) for all the various states, and which could be seen as the reflection of emitting electrons distribution. However, the break in BPL during state Q1 and F1 is different from all the other states which probably suggests the involvement of different emission regions.

\begin{table*}
 \centering
 \caption{Results of $\gamma$-ray SED analysis for various observed states.}
 \begin{tabular}{c c c c c c c c}
 \hline \
 Various & F$_{0.1-300 \rm{GeV}}$& Luminosity & PowerLaw & & & -log(Likelihood) & TS$_{curve}$ \\
 states & (10$^{-7}$ ph cm$^{-2}$ s$^{-1}$) & (10$^{48}$ erg s$^{-1}$) & $\Gamma$ & & & & \\
  \hline
 Q1       & 1.42$\pm$0.18& 0.15    & -2.27$\pm$0.09 & -- & -- & 24756.95  & -- \\ 
 F1       & 3.46$\pm$0.21& 0.60    & -2.02$\pm$0.04 & -- & -- & 40874.32 & --\\
 \hline
 Preflare & 1.61$\pm$0.08&0.25 & -2.20$\pm$0.03 & -- & -- & 207593.05 & --\\
 P1       & 3.87$\pm$0.42&0.80 & -1.85$\pm$0.07 & -- & -- & 11168.86  & --\\
 P2       & 4.72$\pm$0.27&0.93 & -1.95$\pm$0.04 & -- & -- & 32409.23 & --\\
 P3       & 7.02$\pm$0.35&1.04 & -2.18$\pm$0.04 & -- & -- & 35172.62 & --\\
 P4       &12.00$\pm$0.28&2.92& -1.98$\pm$0.02 & -- & -- & 103977.92& -- \\
\hline
    && & LogParabola & \\
    & && $\alpha$ & $\beta$ &  & & \\
  \hline 
  Q1     & 1.14$\pm$0.20& 0.14    & 2.25$\pm$0.13 & 0.25$\pm$0.11 & -- & 24752.63  & 8.64 \\
  F1     & 3.30$\pm$0.22&  0.60   & 1.98$\pm$0.05 & 0.06$\pm$0.03 & -- & 40868.08 & 12.48 \\
  \hline
Preflare & 1.50$\pm$0.09&0.22 & 2.18$\pm$0.07 & 0.04$\pm$0.03 & -- & 207588.33 & 9.44 \\
 P1      & 3.37$\pm$0.42&0.81 & 1.70$\pm$0.10 & 0.12$\pm$0.05 & -- & 11165.35  & 7.02 \\
 P2      & 4.37$\pm$0.28&0.84 & 1.74$\pm$0.08 & 0.08$\pm$0.03 & -- & 32400.11 & 18.24 \\
 P3      & 6.77$\pm$0.36&0.96 & 2.06$\pm$0.07 & 0.07$\pm$0.03 & -- & 35169.75 & 5.74 \\
 P4      &11.20$\pm$0.32&2.45 & 1.92$\pm$0.02 & 0.06$\pm$0.01 & -- & 103956.66& 42.52 \\  
 \hline \ 
   & && PLExpCutoff & E$_{cutoff}$ & & & \\
   & && $\Gamma_{PLEC}$ & [GeV] & && \\
\hline   
 Q1      & 1.18$\pm$0.20& 0.14    & -1.71$\pm$0.27 & 1.73$\pm$0.91  & -- & 24752.13  & 9.64 \\
 F1      & 3.31$\pm$0.23&  0.60   & -1.92$\pm$0.07 & 22.02$\pm$13.22& -- & 22397.94 & 36952.76\\
 \hline
Preflare & 1.57$\pm$0.08&0.24 & -2.14$\pm$0.04 & 29.72$\pm$5.63 & -- & 135578.86 & 144028.38 \\
 P1      & 3.50$\pm$0.42&0.85 & -1.60$\pm$0.13 & 9.77$\pm$5.32  & -- & 11164.87  & 
 7.98 \\
 P2      & 4.89$\pm$0.28&1.20   & -1.88$\pm$0.06 & 19.17$\pm$9.70 & -- & 32399.41 & 19.64 \\
 P3      & 6.90$\pm$0.36&1.00  & -2.11$\pm$0.06 & 23.80$\pm$16.81& -- & 35170.84 & 3.56\\
 P4      &11.50$\pm$0.29&2.54 &-1.89$\pm$0.02 & 28.09$\pm$6.46 & -- & 77860.30& 52235.24 \\   
 \hline
   && & Broken PowerLaw & & E$_{break}$& &  \\
   & && $\Gamma_1$ & $\Gamma_2$ & [GeV] & & \\
 \hline
 Q1       & 1.23$\pm$0.19& 0.15    & -2.01$\pm$0.14 & -4.27$\pm$1.08 & 1.88$\pm$0.09 & 24751.87  & 10.16 \\
 F1       & 3.35$\pm$0.22&  0.59   & -1.94$\pm$0.06 & -2.27$\pm$0.15 & 1.99$\pm$0.00 & 40867.83 & 12.98 \\
 \hline
 Preflare & 1.47$\pm$0.25&0.21 & -1.93$\pm$0.25 & -2.48$\pm$0.12 & 0.70$\pm$0.17 & 207584.79 & 16.52 \\
 P1       & 3.52$\pm$0.44&1.03 & -1.60$\pm$0.19 & -2.09$\pm$0.18 & 1.10$\pm$0.75 & 11166.94  & 3.84 \\
 P2       & 4.43$\pm$0.42&1.08 & -1.75$\pm$0.14 & -2.15$\pm$0.11 & 0.89$\pm$0.29 & 32401.65 & 15.16 \\
 P3       & 6.86$\pm$0.36&1.16 & -2.09$\pm$0.09 & -2.32$\pm$0.15 & 0.90$\pm$0.95 & 35172.32 & 0.60 \\
 P4       & 11.30$\pm$0.31&3.02& -1.82$\pm$0.04 & -2.17$\pm$0.05 & 1.00$\pm$0.22 & 103959.84& 36.16 \\
 \hline 
 \end{tabular}
\end{table*}
\subsection{Correlations Studies}
To investigate correlation, if any, between flux variations in the $\gamma$-ray band and other wavelengths, we looked into the archives for the availability of data at multiple wavelengths. We could find data only in the radio band at 15 GHz that overlaps with the $\gamma$-ray light curve. The site of $\gamma$-ray production in blazars is controversial. Some study suggests that $\gamma$-rays are generally produced at the sub-parsec scale from the apex of the jet (\citealt{Dermer_1994}, \citealt{Blandford_1995}, \citealt{Ghisellini_1996}). The radio core, which is the source of radio emission in the jet represents a transition zone from synchrotron self absorbed region to optically thin region \citep{Blandford_1979}. It is also believed that the radio core is generally located at pc scales from the central super massive black hole (SMBH). Correlation between $\gamma$ and radio light curves will help one to constrain the location of these emission regions. The long term radio and $\gamma$-ray light curves are shown in Figure 7. From Figure 7, it is evident that flux variations in the $\gamma$-ray band is more compared to the radio band that suggest that $\gamma$-rays are produced close to the base of the jet, while, the radio emission is produced at much farther distance from the SMBH. We correlated the $\gamma$-ray and radio light curves for a bin size of 20 days using the discrete correlation function method of \citet{Edelson_1988}. The correlation function is shown in Figure 8. We found a strong correlation with the correlation coefficient greater than 70\% with the radio emission leading the $\gamma$-ray emission by 70 days. This suggests that the $\gamma$-ray and radio emissions are produced at different locations along the jet axis. Similar result on radio leading the $\gamma$-ray flux variations is also seen in the blazar 3C 84, where \citet{Britzen_2019} found the radio emission to lead the $\gamma$-ray emission by 300-400 days. From the observed time lag, following \citet{Prince_2019} we estimated the separation between $\gamma$-ray and radio emission as ~11 pc along the jet axis for an average $\beta_{app}$ value of 13.98 and a $\theta$ of 4.2$^{\circ}$ (\citealt{Liodakis_2017}).

We also estimated the significance of the DCF peaks observed in cross-correlation. For that, we simulated 1000 $\gamma$-ray light curves by following the method mentioned in \citet{Emmanoulopoulos_2013} and incorporated into a code by \citep{Connolly_2016} and available for use\footnote{https://github.com/samconnolly/DELightcurveSimulation}. For simulating the
$\gamma$-ray light curves we used power law as the shape
of the power spectral densities (PSD; slope = 1.5) and assumed a lognormal form for the probability density function. This agrees well with the observations as the observered flux distribution also has a lognormal shape. We also simulated the radio light curve with PSD power law slope 2.0 as suggested by \citet{Max-Moerbeck_2014} for the significance estimation.
Further, the simulated $\gamma$-ray light curves are cross-correlated with the simulated radio light curve. A 2$\sigma$ and 3$\sigma$ significance for each time lag was estimated and plotted horizontally in cyan and greenyellow color in Figure 8. 

\begin{figure}
%\centering
 \includegraphics[scale=0.43]{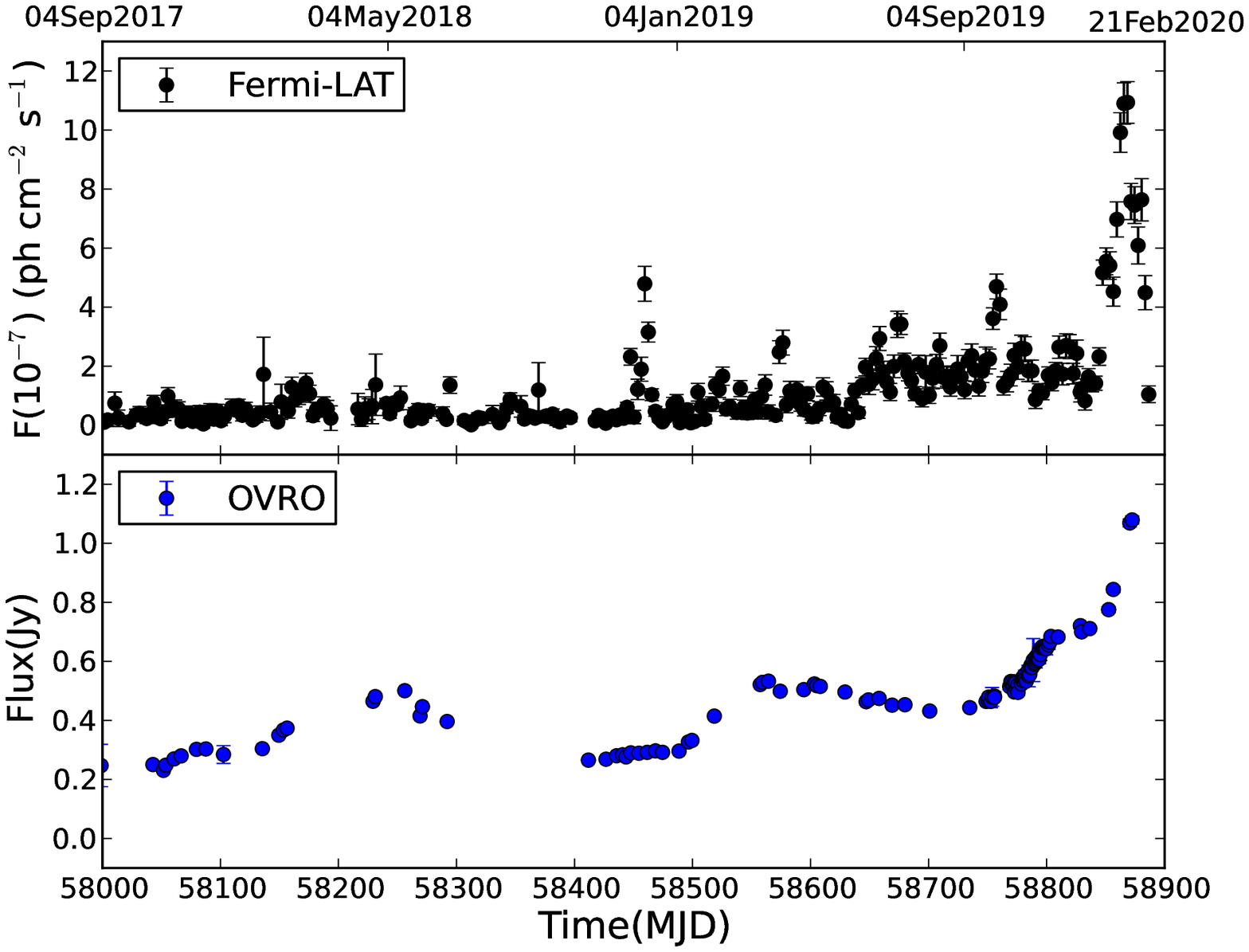} 
 \caption{The long term $\gamma$-ray and radio light curve used for correlation study.}
 \includegraphics[scale=0.43]{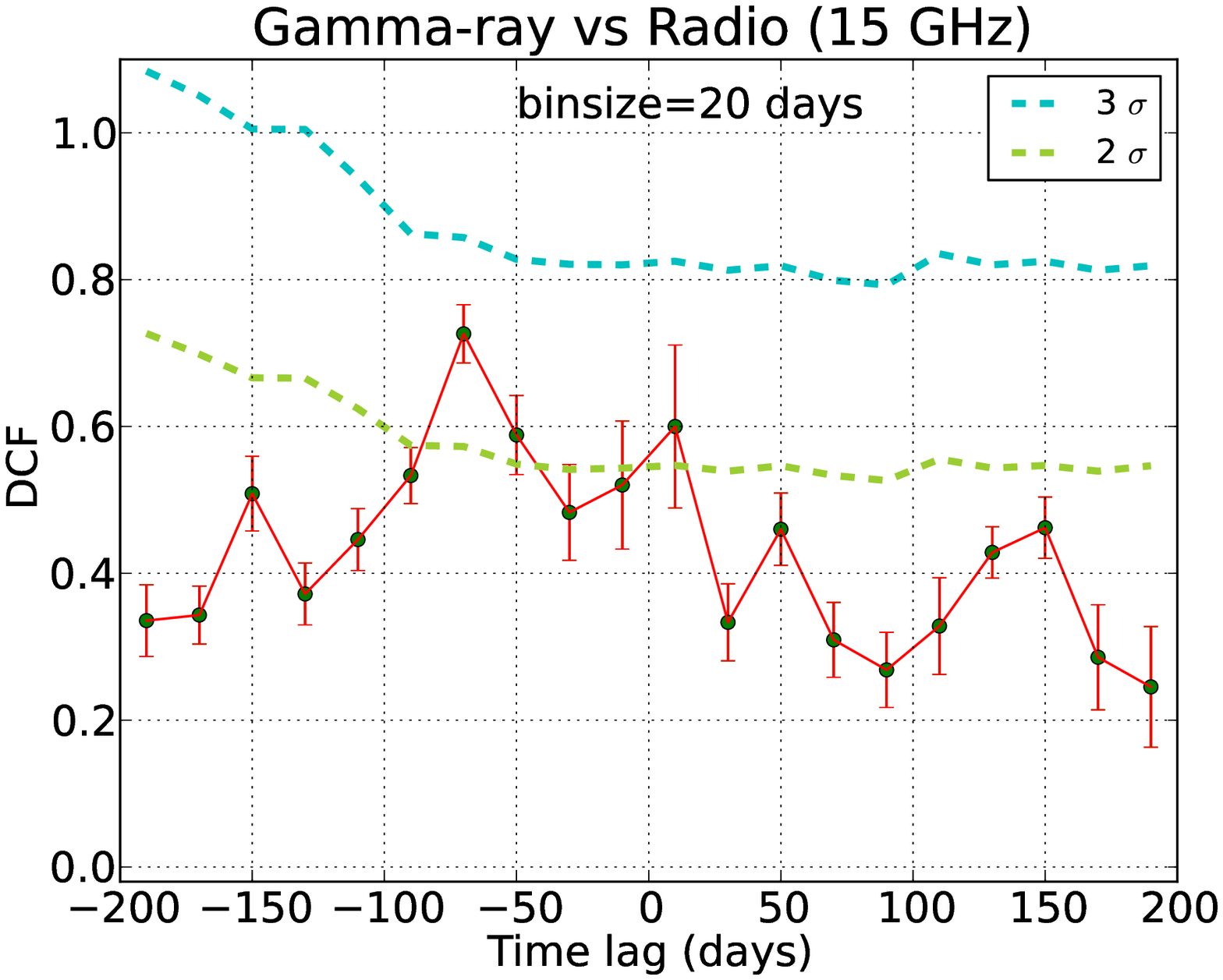} 
 \caption{Cross-correlation of $\gamma$-ray and radio emission.}
\end{figure}

\subsection{Multi-wavelength light curves}
We also looked for the availability of multi-wavelength (MW) data for this source along with the {\it Fermi}-LAT, but apparently, no multi-wavelength observations are available for the pre-flare and flare state except state P4. Therefore, we collected the observations from the Swift-XRT/UVOT and analyzed the X-ray and UV/optical data. Though we do not have a good number of observations in X-ray and UV/optical, the corresponding multi-wavelength light curve for P4 is shown in Figure 9. The significant peaks observed in $\gamma$-ray around MJD 58865 seem missing in X-ray and UV/optical because of the unavailability of observational data. The observation before MJD 58865 in X-ray and UV/optical light curve seems to peak around MJD 58865, and the observation after MJD 58865 suggests the decay in the flux after MJD 58865. Because of the highly sparsed data in X-ray and UV/optical, we could not do correlatio and the time variability studies. 

We calculated the fastest $\gamma$-ray variability time scale during the flaring states of the source using
\begin{equation}
\centering
%t_{var} =  \ln 2 \frac{t_2 - t_1}{\ln(F_2 / F_1)}
F_2 = F_1. 2^{{(t_2-t_1)}/\tau_d}
\end{equation}
where F$_{1}$ and F$_{2}$ are the fluxes at consecutive times t$_1$ and t$_2$ respectively and $\tau_d$ is the flux doubling time also known as variability time.
We found the fastest variability time during P4 with a value of 9.01$\pm$0.78 hr between MJD 58874.25 and 58874.75, during which time the fluxes are 3.95 and 9.95 ($\times10^{-7}$ ph cm$^{-2}$ s$^{-1}$) respectively. 
\begin{figure}
 \includegraphics[scale=0.43]{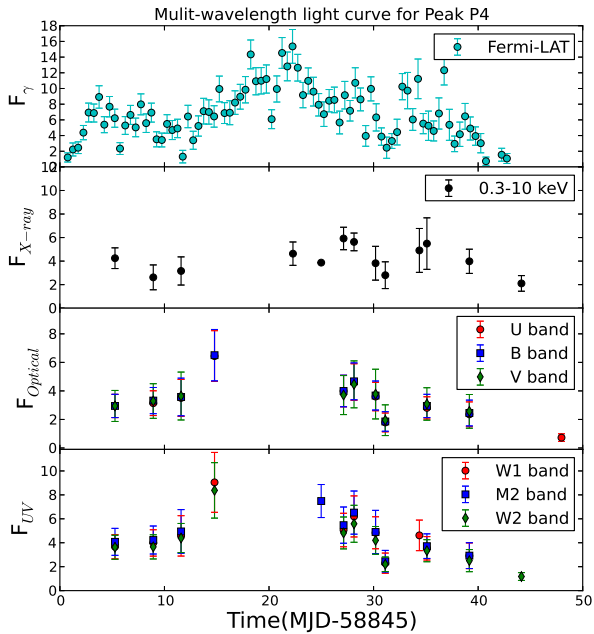}
 \caption{Multi-wavelength light curves of P4. F$_{\gamma}$ is in units of 10$^{-7}$ph cm$^{-2}$ s$^{-1}$, F$_{X-ray}$ has unit of 10$^{-12}$ erg cm$^{-2}$ s$^{-1}$, and F$_{optical}$ and F$_{UV}$ is in units of 10$^{-11}$ erg cm$^{-2}$ s$^{-1}$.}
\end{figure}

We also looked for the {\it Swift}-XRT/UVOT observations for the state Q1 and F1, and we found only few observations (hence the multi-wavelength light curvs are not shown for Q1 and F1). The X-ray and optical/UV SED was prepared from that for both the states and used in the final MW spectral energy distribution (SED) modeling in the next section. 

\section{Results: Broadband SED Modeling}
Fitting the observed SED of the source during various activity states is an ideal way to constrain the physical processes responsible for the broadband emission. Towards this, we looked at the availability of multiband data covering UV, optical and X-rays during the various activity periods of the source marked in Figure 1. Multi-wavelength data was sparsely available only for Q, F1 and  P4 states. The multi-wavelength light curves for the state P4 is shown in Figure 9. Here too, X-ray and UV/optical observations are not available during the peak of the $\gamma$-ray flares. 
We modelled the observed SEDs during Q, F1 and P4 states using the {\it python} based and publicly available time dependent model GAMERA\footnote{http://joachimhahn.github.io/GAMERA} (\citealt{Hahn_2015}).
 
Among all the flares observed in 2019 and 2020 (state F2), only P4 has good coverage of simultaneous multi-wavelength data and hence provides an ideal situation to go for SED modeling. The other states like Q1 and F1 have very few observations in X-ray and optical/UV. The MW SEDs for states Q1 and F1 are also produced and their parameters are compared with the brightest state P4.
GAMERA calculates the propagated electron spectrum for a given initial injected electron spectrum by solving the transport equation given in equation (3), and further estimates the synchrotron, synchrotron self-Compton (SSC), and inverse Compton (IC) emissions.

\begin{equation}\label{8}
\frac{\partial N(E,t)}{\partial t}=Q(E,t)-\frac{\partial}{\partial E}\Big(b(E,t) N(E,t)\Big)
\end{equation}
where, $N(E,t)$ is the propagated electron spectrum estimated at a time 't' for the initially injected electron spectrum ($Q(E,t)$). $b(E,t)$ is dedicated to the radiative loss caused by physical processes, viz. synchrotron, SSC, and IC scattering. For the model fits here, we considered a single zone emission model with a log parabola electron distribution.

\subsection{Jet Parameters}
Further, we tried to derive few jet parameters for this source based on our observational results.  
We can constrain the Doppler factor from the $ \gamma $-$ \gamma $ opacity argument (\citealt{Dondi_1995}, \citealt{Ackermann_2010}) and which can be derived numerically by using the highest energy photon detected in $\gamma$-ray during the flare. The argument says that if the $ \gamma $-$ \gamma $ interaction optical depth for the high energy photon is one, the minimum Doppler factor can be defined as:
\begin{equation}
 \delta_{min} \cong \left[ \frac{\sigma_T d_L^2 (1+z)^2 f_x \epsilon}{4 t_{var} m_e c^4}  \right] ^{1/6}
\end{equation}
where, $\sigma_T$ is the Thompson scattering cross section (6.65$\times$10$^{-25}$ cm$^2$), d$_L$ is the luminosity distance (4.175 Gpc), f$_x$ is the X-ray flux measured in the 0.3$-$10 keV (4.908$\times$10$^{-12}$ erg/cm$^2$/s), $\epsilon$ = E/$m_e c^2$, E is the highest photon energy ($\sim$77 GeV), and t$_{var}$ is the the observed variability time (9.01$\pm$0.78 hr). All the observed values are measured during the flare and are contemporaneous in nature. The Doppler factor derived from equation (4) is $\sim$12.
In general, blazar has a similar bulk Lorentz and Doppler factor for the emitting blob, i.e., $ \Gamma $ $\sim$ $ \delta $ and which can provide the upper limit on the viewing angle of the jet, $\theta$ $\leq$ 1/$\delta_{min}$ = 4.8$^{\circ}$.
The size and the location of the emission region can be constrained for the flaring period. The upper limit on size can be derived from the causality relation R  $\sim$ c t$_{var}$ $\delta_{min}$/(1+z) $\sim$ 6.94$\times$10$^{15}$ cm. Assuming a conical jet scenario where the emission is produced across the entire jet area suggests the flaring site close to the central engine and the distance can be estimated as d $\sim$ 2 c t$_{var}$ $\delta_{min}^2$/(1+z) $\sim$ 0.05 pc \citet{Abdo_2011}.

The total isotropic $\gamma$-ray luminosity can also be estimated for all the spectral shapes (PL, LP, PLEC, and BPL) by following the relation,
\begin{equation}
 L_{\gamma} = 4 \pi D_L^2 \int_{E_{min}}^{E_{max}} E \frac{dN}{dE} dE
\end{equation}
where E$_{min}$ is the lower energy range of {\it Fermi}-LAT (i.e. 100 MeV) and E$_{max}$ is the energy of the highest photon detected during a particular period, dN/dE represents the various spectral models and D$_L$ is the luminosity distance (4.259 Gpc). The $\gamma$-ray luminosity corresponding to each spectral models in their various states are shown in Table 4. The obtained values suggest that the $\gamma$-ray luminosity is more during the higher states P1, P2, P3, and P4 compared to the Q1 and F1 states.

\subsection{External seed photons}
In the leptonic scenario, BLR is believed to be the main source of seed photons that get up-scatted in the jet through the IC scattering and produces the high energy peak of the SED. Considering BLR as a thin spherical shell, makes easy to estimate the size of the BLR (\citealt{Ghisellini_2009}) and which can be scaled as, R$_{BLR}$ = 10$^{17}$ L$_{d,45}^{1/2}$, where L$_{d,45}$ is the disk luminosity in units of 10$^{45}$ erg s$^{-1}$. The disk luminosity and mass of the SMBH for this source are estimated by \citet{Brotherton_2015}, and the values are M$_{SMBH}$ = 3.98$\times$10$^{8}$ M$_{\odot}$ and L$_{disk}$ = 9.2$\times$10$^{45}$ erg s$^{-1}$ after bolometric correction from \citet{Netzer_2019}. After using the above value of L$_{disk}$ the R$_{BLR}$ is derived as 3.03$\times$10$^{17}$ cm $\sim$ 0.1 pc. Comparing the R$_{BLR}$ with the location of the emission region ($\sim$ 0.05 pc), we conclude that the emission region is located at the inner boundary of the BLR. In our SED modeling we considered that the emission site is located within the BLR and the BLR plays an important role to produce the high energy peak of the SED. For our modeling purpose we estimated the BLR photon energy density in the comoving frame by,
\begin{equation}
 U_{BLR}^{'} = \frac{\Gamma^2 \eta_{BLR} L_{disk}}{4 \pi c R^2_{BLR}}
\end{equation}
where the $\eta_{BLR}$ is the fraction of disk emission processed in BLR, and typically it is around 10\%, and c is the speed of light in vacuum. 

The contribution of direct disk emission as seed photons for IC scattering can not be ignored and hence we also estimated the accretion disk photon energy density in the comoving frame from \citet{Dermer_2009},
\begin{equation}\label{9}
U'_{disk}=\frac{0.207 R_g l_{Edd} L_{Edd}}{\pi c z^3 \Gamma^2}
\end{equation}
where, R$_g$, l$_{Edd}$ = L$_{disk}$/L$_{Edd}$, and $z$ are the gravitational radius, the Eddington ratio, and the location of the emission site from the SMBH respectively. The 
gravitational radius is found to be R$_g$ = 5.87$\times$10$^{13}$ cm for the black hole mass 3.98$\times$10$^{8}$ M$_{\odot}$. 
We did not consider dusty torus as a source of external photon
field since there is no observational evidence. However, the dusty torus contribution could be important if the emission site is outside the BLR.

The calculated photon energy density in BLR and disk along with the BLR temperature (10$^{4}$K; \citealt{Peterson_2006}) and disk temperature (1.1$\times$10$^{5}$K; estimated from Eddington ratio and BH mass; \citealt{Panda_2018}) were fixed as inputs in GAMERA, and the
parameters of the input injected electron distributions were kept free while modeling the multi-wavelength SED. The size of the emitting blob was fixed from the variability time calculation, and the jet magnetic field was set free to obtain the good fit value of the multi-wavelength SED. 
\begin{figure*}
 \includegraphics[scale=0.45]{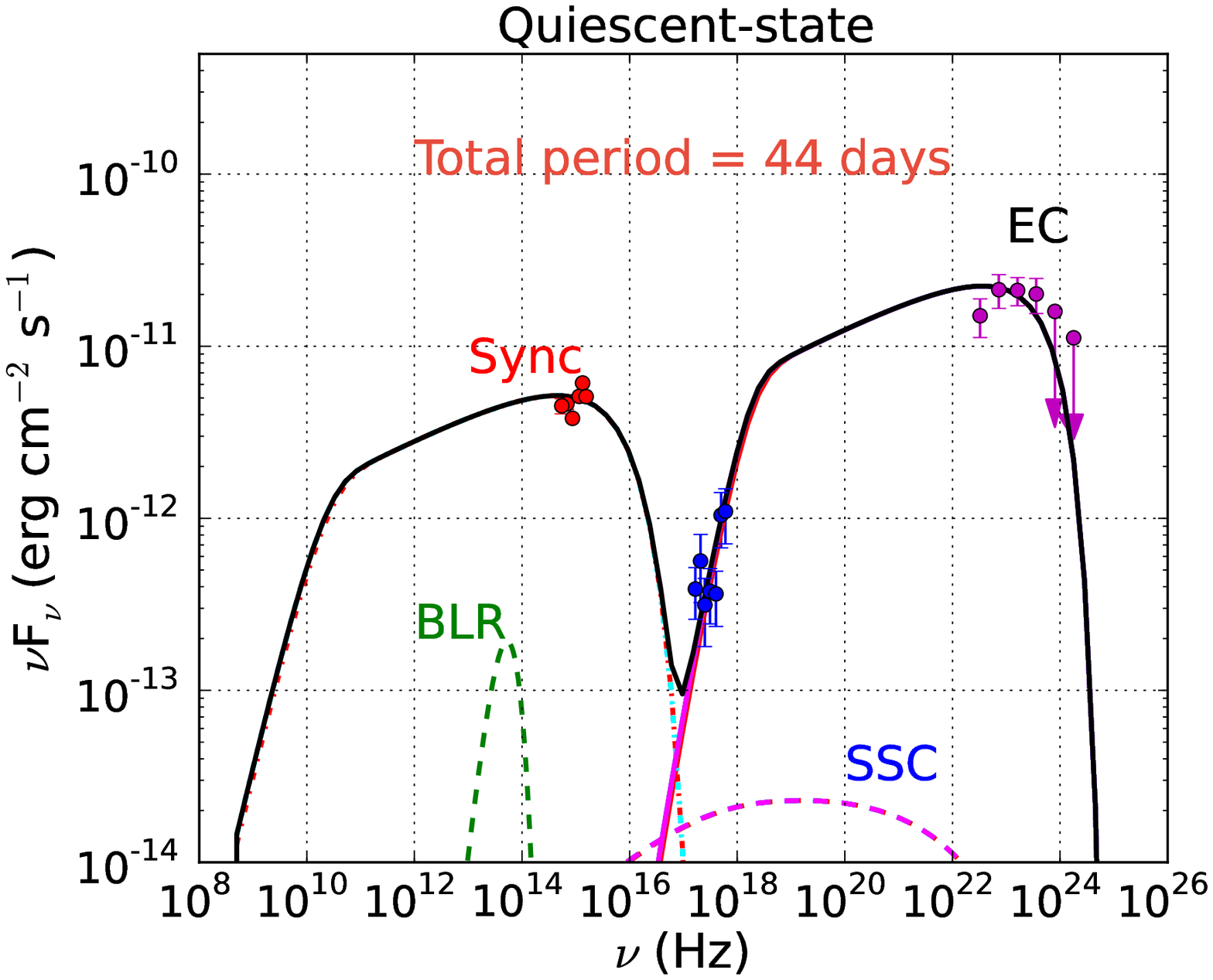}
 \includegraphics[scale=0.45]{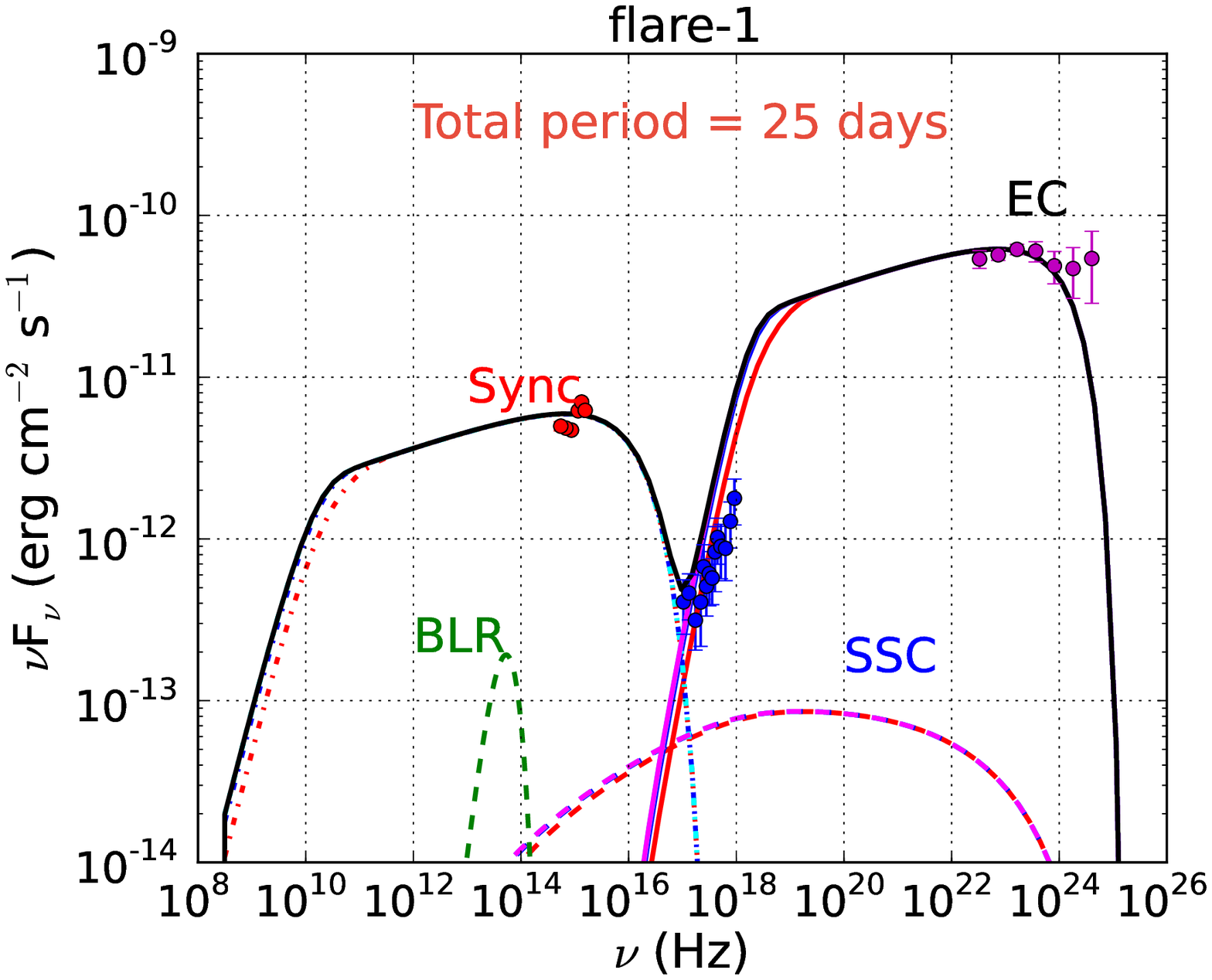}
 \includegraphics[scale=0.45]{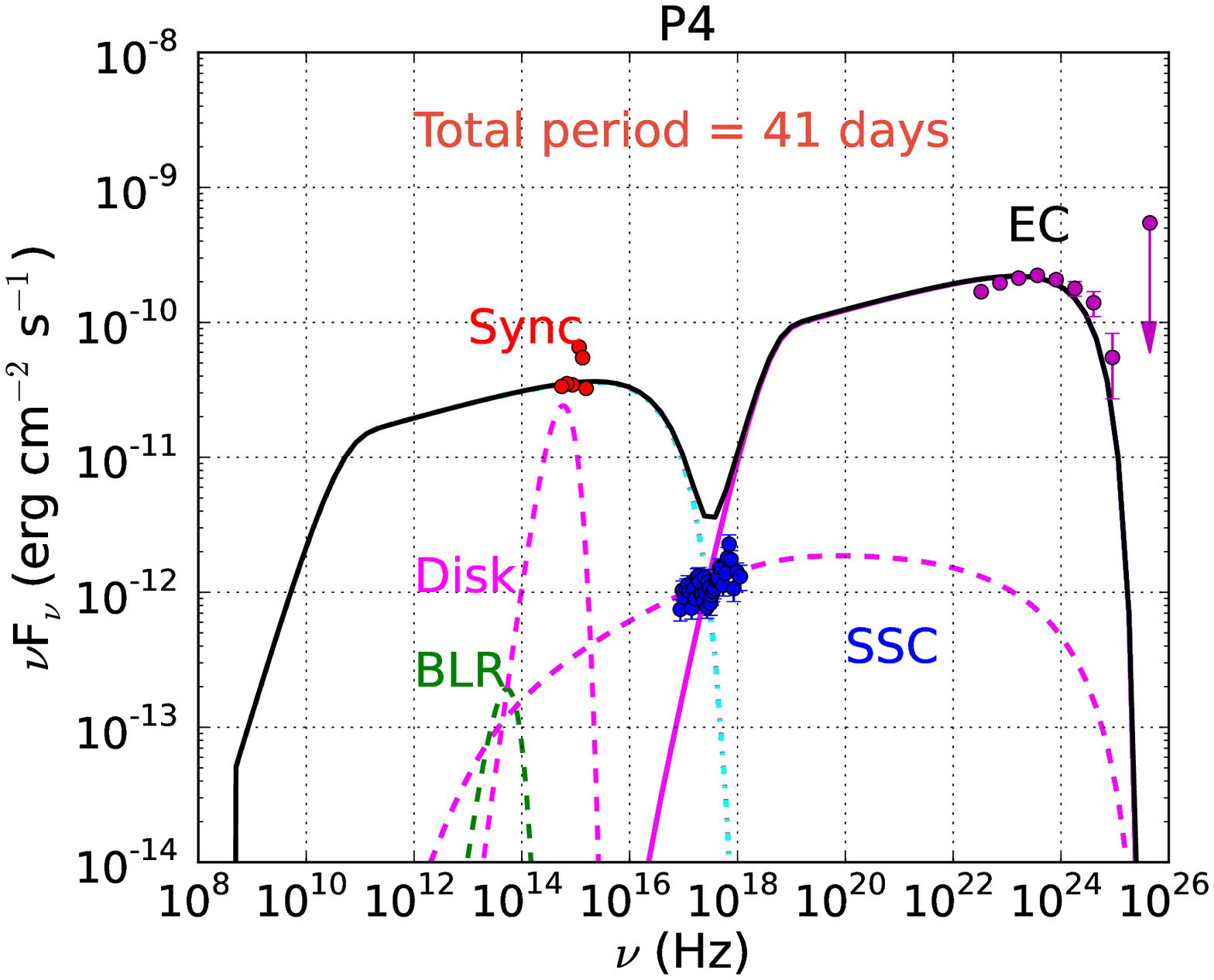}
 \caption{Multi-wavelength SED of states Q1, F1 and the high state P4.}
\end{figure*}

\subsection{Modeling results}
We carried out the SED modeling for the states Q1, F1 and the high state P4, and the multi-wavelength SED modeling plots are shown in Figure 10. The best fit model parameters for SED modeling are shown in Table 5. The low energy peak is successfully constrained by the Synchrotron process and the $\gamma$-ray high energy peak with the IC process. The SSC mechanism well describes the X-ray emission in the high state P4, whereas in Q1 and F1 it is explained by external Compton process. The size of the emission region is found to be a bit more from the modeling than the estimated value from the variability time. The magnetic field in the blob is very much similar to the other FSRQ type blazars like PKS 1510$-$089 \citep{Prince_2019a}, 3C 279 \citep{Prince_2020}, 3C 454.3 \citep{Das_2020} etc. The Doppler factor and the Lorentz factor were optimized to 20 and 15.5 to obtain the best fit to the multi-wavelength SED.

\begin{table*}
\centering
\caption{Multi-wavelength SED modeling results with the best fitted parameters values. The input injected electron distribution is LogParabola with reference energy 60 MeV. }
 \begin{tabular}{c c c c c}
 \hline
 high state& Parameters & Symbols & Values & Period \\
 \hline
 &BLR photon density& U$^{'}_{BLR}$ & 3.83 erg/cm$^3$ & \\
 &BLR temperature & T$^{'}_{BLR}$ & 1.0$\times$10$^4$ K & \\
 &Disk photon density&U$^{'}_{disk}$ & 2.21$\times$10$^{-6}$ erg/cm$^3$ & \\
 &Disk temperature & T$^{'}_{disk}$ & 1.0$\times$10$^6$ K & \\
 & Size of the emitting zone& R & 2.6$\times$10$^{16}$ cm & \\
 & Doppler factor of the emitting zone& $\delta$ & 20.0 & \\
 & Lorentz factor of the emitting zone& $\Gamma$ & 15.5 & \\
 \hline
Q1 & &&& 44 days\\ 
 & Min Lorentz factor of emitting electrons & $\gamma_{min}$& 5.0 &\\
 & Max Lorentz factor of emitting electrons & $\gamma_{max}$& 6.3$\times$10$^{3}$ &\\
 & Input injected electron spectrum (LP) & $\alpha$ & 1.68 & \\
 & Curvature parameter of the PL spectrum & $\beta$& 0.005 & \\
 & Magnetic field in emitting zone & B & 4.7 G & \\
 & Jet power in electrons & P$_{j,e}$ & 4.66$\times$10$^{44}$ erg/s & \\
 & Jet power in magnetic field & P$_{j,B}$ & 1.29$\times$10$^{46}$ erg/s & \\
 & Jet power in protons & P$_{j,P}$ & 4.35$\times$10$^{44}$ erg/s& \\
 & Total jet power & P$_{jet}$ & 1.37$\times$10$^{46}$ erg/s& \\
 \hline
F1 & &&& 25 days\\ 
 & Min Lorentz factor of emitting electrons & $\gamma_{min}$& 5.0 &\\
 & Max Lorentz factor of emitting electrons & $\gamma_{max}$& 1.1$\times$10$^{4}$ &\\
 & Input injected electron spectrum (LP) & $\alpha$ & 1.77 & \\
 & Curvature parameter of the PL spectrum & $\beta$& 0.005 & \\
 & Magnetic field in emitting zone & B & 3.0 G & \\
 & Jet power in electrons & P$_{j,e}$ & 1.30$\times$10$^{45}$ erg/s & \\
 & Jet power in magnetic field & P$_{j,B}$ & 5.48$\times$10$^{45}$ erg/s & \\
 & Jet power in protons & P$_{j,P}$ & 1.20$\times$10$^{45}$ erg/s& \\
 & Total jet power & P$_{jet}$ & 7.98$\times$10$^{45}$ erg/s& \\ 
 \hline
P4 & &&& 41 days\\
 & Min Lorentz factor of emitting electrons & $\gamma_{min}$& 8.0 &\\
 & Max Lorentz factor of emitting electrons & $\gamma_{max}$& 1.6$\times$10$^{4}$ &\\
 & Input injected electron spectrum (LP) & $\alpha$ & 1.77 & \\
 & Curvature parameter of the PL spectrum & $\beta$& 0.005 & \\
 & Magnetic field in emitting zone & B & 3.9 G & \\
 & Jet power in electrons & P$_{j,e}$ & 4.84$\times$10$^{45}$ erg/s & \\
 & Jet power in magnetic field & P$_{j,B}$ & 9.26$\times$10$^{45}$ erg/s & \\
 & Jet power in protons & P$_{j,P}$ & 3.43$\times$10$^{45}$ erg/s& \\
 & Total jet power & P$_{jet}$ & 1.75$\times$10$^{46}$ erg/s& \\
\hline
 \end{tabular}

\end{table*}

We also used the derived parameters to estimate the individual jet power in electrons, magnetic field, and protons. The total jet power can be defined as,
\begin{equation}
 P_{jet} = \pi r^2 \Gamma^2 c (U_e + U_B + U_P)
\end{equation}
where, U$_e$, U$_B$, and U$_P$ are the energy density in electrons, magnetic field, and protons. The size of the emitting zone and its Lorentz factor is denoted by $r$ and $\Gamma$. The jet is considered as plasma of leptons and protons with the ratio of 20:1.
The total jet power calculated here is always lesser than the total Eddington luminosity of the source. The power calculated for individual components are mentioned in the Table 5. Comparing with the other flaring blazars like PKS 1510$-$089 \citep{Prince_2019a}, 3C 279 \citep{Prince_2020}, and 3C 454.3 \citep{Das_2020} the total and individual components powers are in good agreement. A blazar sample has been studied by \citet{Ghisellini_2015}, and they found that most of the blazars in their sample have L$_{disk}$/L$_{Edd}$ = 0.1 and using this ratio the Eddington luminosity estimated in our case is 9.2$\times$10$^{46}$ erg/s, which is much greater than the total jet power estimated here. As we can see in Table 5, the total jet power is dominated by the magnetic field and hence powers the blazar, whereas both the leptons and protons power do not provide sufficient power and are unable to supply the energy to the radio lobes. Comparing the values obtained during various states suggest that more electrons and protons power are needed to transit the source from state Q1 to F1 and P4. Surprisingly the magnetic field and magnetic power obtained during state Q1 is more compared to state F1 and P4. A large value of minimum and maximum energy is required in electrons to make the source transit from low state (Q1) to high state (F1 and P4). 

%##############################
\section{Results: Flux distribution of OQ 334}
%####################### New
Analysis of the $\gamma$-ray light curves, suggest that many blazars show lognormal behaviour in their flux distributions \citep{Ackermann_2015, 2018RAA....18..141S, 2018Galax...6..135R}. Among blazars, such lognormal behaviour was first detected in BL Lac, from RXTE observation \citep{2009A&A...503..797G}, later it was observed in many blazars at various energy bands and time scales \citep{2010A&A...520A..83H, 2010A&A...524A..48T,pankaj_ln, my421, my1011, 2018MNRAS.480L.116S, 2020MNRAS.491.1934K}. Lognormal flux distributions were initially found in the X-ray emission of black hole binary Cygnus X-1 \citep{lognorm_xrb,2002A&A...385..377Q, 2009A&A...503..797G}, and are generally explained by the fluctuations in the accretion disc, which imply multiplicative processes \citep{2005MNRAS.359..345U, 2010LNP...794..203M}. However, fast (minute time scale) variability in the blazars' lightcurves, is difficult to produce in the disc \citep{narayan}, and supports to originate in the jet.
 \citet{Biteau} have shown that log-normal flux distribution can be explained by the multiplicative processes, however, according to \citet{2020ApJ...895...90S}
log-normality of measured flux values need not imply multiplicative process.
On the other hand, Gaussian perturbation in the particle acceleration time scale is capable of producing a lognormal flux distribution \citep{2018MNRAS.480L.116S}. 

We studied the flux distribution property for the source OQ334, using three days binned $\gamma$-ray flux lightcurve. To select lightcurve with good statistics, we considered flux points for which TS>= 9, and also the flux points detected at greater than 2-$\sigma$ level, such that $\frac{F}{\Delta{F}}$ > 2. We performed Anderson-Darling (AD) test, where null hypothesis probability value (p-value) < 0.01 would indicate non-Gaussianity of the data. AD test results show that the test statistics (r) and p-value for flux in linear scale are 15.06 and $1.5\times10^{-3}$, while r-value and p-value for the flux in log-scale are 0.57 and 0.13 respectively, which implies that the flux distribution is lognormal. To quantify this, we further fit the normalized histogram of the logarithm of flux with the Gaussian and lognormal probability density functions (PDFs) (Figure 11), these PDFs are shown by \citet{2018RAA....18..141S}. We found that the lognormal PDF significantly fits the distribution better with reduced chi-square, ${\chi^{2}}_{red} {\approx}$ 0.93 for 8 degrees of freedom (dof), than the Gaussian PDF (${\chi^{2}}_{red} {\approx}$ 2.80 for 8 dof). 
 We further investigated the linear dependence of the average flux on its excess variance, which is an important feature for lognormal behaviour. For that purpose, we considered the Poisson noise corrected excess variance as given by, $\sigma_{XS}= \sqrt{S^2-\overline{\sigma_{err}^2}}$ ; where $S^2$ is the sample variance and 
${\overline{\sigma_{err}^2}}$ represents the mean of the square of the measurement errors \citep{Vaughan} .
%The linear dependence of the average flux on its excess variance, $\sigma_{XS}= \sqrt{S^2-\overline{\sigma_{err}^2}}$ is an important feature for lognormal behaviour.
The flux-rms plot is shown in Figure 12, where data is binned for a period of 50 days to obtain sufficient statistics. The scatter plot is well fitted by a linear function with slope 0.43$\pm$0.05. Furthermore, we computed the Spearman's rank correlation coefficient (${r_{s}}$) and the correlation probability (prob). The values of ${r_{s}}$ and prob are found as 0.76 and $2.6\times10^{-3}$, indicating a strong correlation between flux and excess variance.  
%\section{Discussion}
The observed lognormal behaviour in the flux distribution and the proportionality between the average amplitude of variability to the flux, suggest that the variation in flux is log-normal. 

%##############################
\begin{figure}
\centering
\includegraphics[scale=1.2]{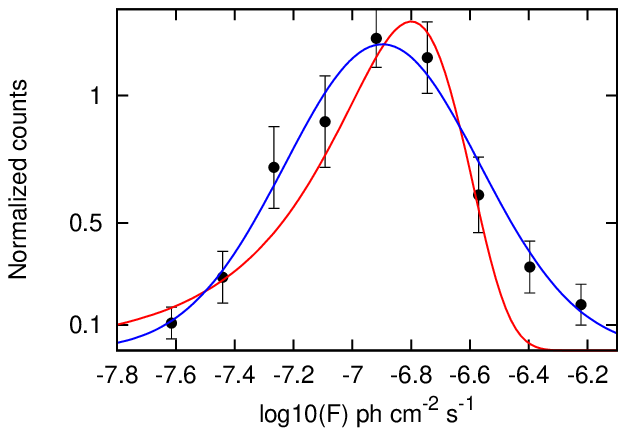}
\caption{Histogram of the logarithm of 3 days binned $\gamma$-ray fluxes. The red and blue lines represent the Gaussian and lognormal PDFs respectively.} 
 
%\label{fig:273_corr}
\end{figure}
\begin{figure}
\centering
\includegraphics[scale=0.5,angle=-90]{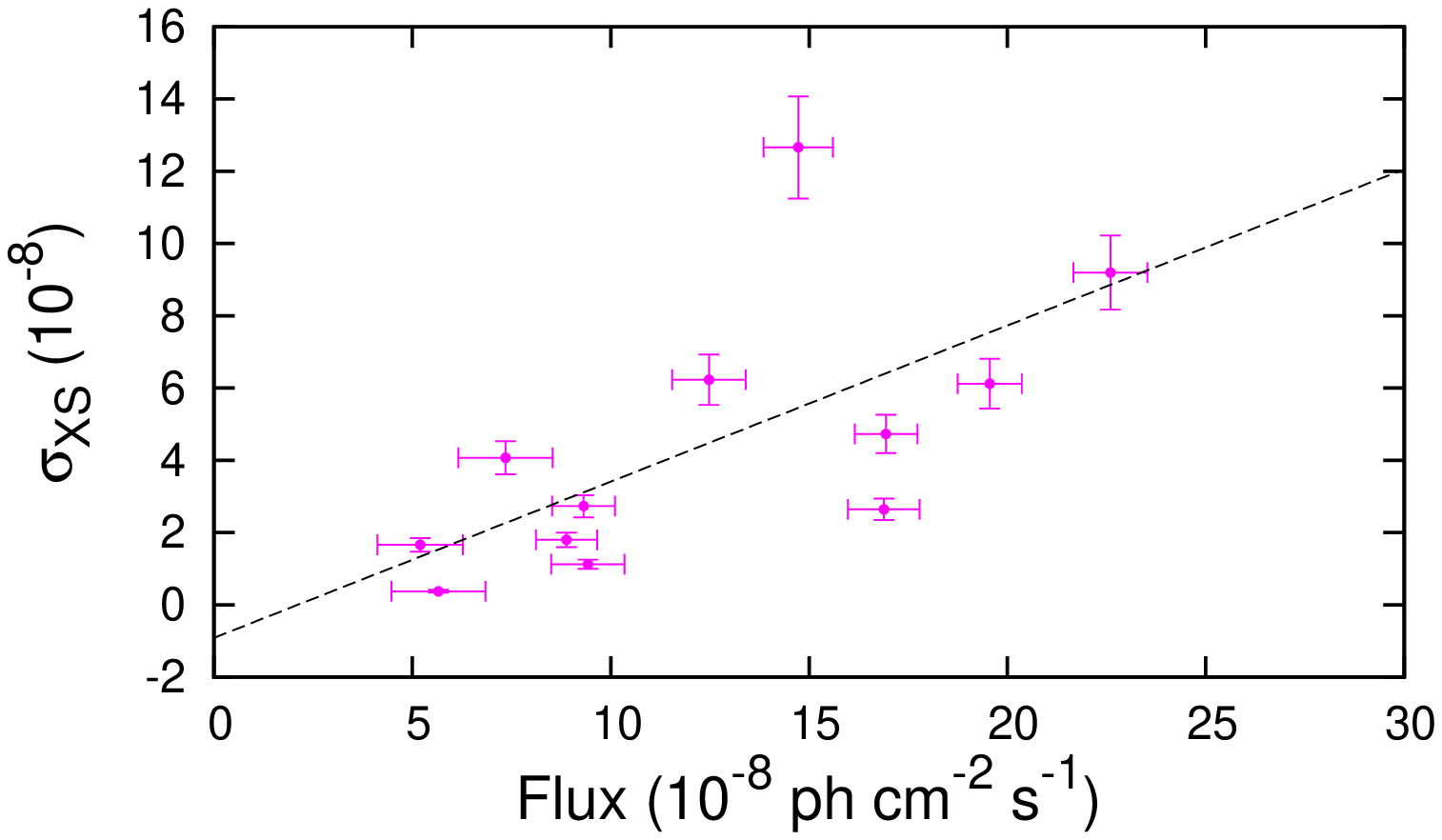}
\caption{Excess variance vs mean flux scatter plot, with the best fit line (black).}
%\label{fig:501_corr}
\end{figure}

\section{Summary}

In this work, we present the first time detailed analysis of the
$\gamma$-ray spectral and temporal
behaviour as well as the broad band SED modelling of the FSRQ OQ 334. We
summarize the main results below,
\begin{enumerate}
 \item The source was faint in the $\gamma$-ray band for about 10 years. It
showed a bright $\gamma$-ray flare
during 2017, returned back to the low brightness state, stayed in the
low brightness state for few months and
again moved to the higher $\gamma$-ray brightness state. Supermimposed on
the high brightness state many small
flares were observed, with the brightest $\gamma$-ray flare occuring in February
2020. Thus the source has shown more
than one episode of flaring activity in the $\gamma$-ray band between 2017
and 2020.
\item During various brightness state of the source such as P1, P2,
P3 and P4 states, no correlation
of the $\gamma$-ray photon index with the total flux of the source was
found.
\item During most of the $\gamma$-ray brightness sates of the source, the
$\gamma$-ray spectrum was well fit by a PLEC spectral model.
\item We found a time lag between the $\gamma$-ray and
radio band light curve at the 2$\sigma$ significance level with the radio variation leading variations
in the $\gamma$-ray band by 70 days. The two emission regions are thus
separated by $\sim$11 pc.
\item We found the fastest variability time on scales of 9.01 $\pm$ 0.78 hrs. Using that we constrained the size and location of the emission region as 6.95$\times$10$^{15}$ cm and 0.05 pc respectively.
\item The broad band SED modelling indicates that the location of the
$\gamma$-ray emission region is inside the BLR.
The observed $\gamma$-ray emission during Q1, F1, and P4 states are the combination of SSC and IC scattering. The physical parameters obtained from SED modelling indicates that more electron and protron power are needed to transit
the source from Q1 to F1 and further in P4 state.
\item The flux distribution shows the lognormal behaviour in the source.
\end{enumerate}

\section*{Acknowledgements}

We thank the referee for providing stimulating comments in order to improve the paper.
The project was partially supported by the Polish Funding Agency National Science Centre, project 2017/26/A/ST9/00756 (MAESTRO 9), and MNiSW grant DIR/WK/2018/12.
R.P thanks Avik Kumar Das for SED modeling discussions. R.P. thanks Prof. Bo$\dot{\rm z}$ena Czerny, Swayamtrupta Panda, and Michal Zajacek for discussions. This work has made use of public {\it Fermi} data obtained from FSSC. This research has also made use of XRT data analysis
software (XRTDAS) developed by ASI science data center, Italy. This research has made use of radio data from OVRO 40-m monitoring programme (Richards et al. 2011) which is supported in part by NASA grants NNX08AW31G, NNX11A043G, and NNX14AQ89G and NSF grants AST-0808050 and AST-1109911.

\section*{Data Availability}
This research has made use of archival data from various sources e.g. {\it Fermi}, {\it Swift}, and {OVRO observatory} and their proper links are given in the manuscript.  All the models and softwares used in this manuscript are also publicly available.

%%%%%%%%%%%%%%%%%%%%%%%%%%%%%%%%%%%%%%%%%%%%%%%%%%



\bibliographystyle{mnras}
\bibliography{reference-list.bib} % if your bibtex file is called example.bib

%\begin{thebibliography}{99}
%\bibitem[Author (2012)]{Author 2012}
%\end{thebibliography}

%%%%%%%%%%%%%%%%%%%%%%%%%%%%%%%%%%%%%%%%%%%%%%%%%%

%%%%%%%%%%%%%%%%% APPENDICES %%%%%%%%%%%%%%%%%%%%%

%\appendix


%If you want to present additional material which would interrupt the flow of the main %%paper,

%%%%%%%%%%%%%%%%%%%%%%%%%%%%%%%%%%%%%%%%%%%%%%%%%%

% Don't change these lines
\bsp	% typesetting comment
\label{lastpage}
\end{document}